\documentclass[journal, onecolumn]{IEEEtran}
\usepackage[utf8]{inputenc}
\usepackage{bbm}
\usepackage{amsmath,amssymb,mathtools}
\usepackage{algorithm,algpseudocode}
\usepackage{hyperref}
\usepackage{booktabs}
\usepackage{thm-restate}
\usepackage{cleveref}
\usepackage{bbold}
\usepackage{bm}
\usepackage{graphicx,psfrag}
\usepackage{pstool}
\usepackage{calc}
\usepackage{comment}
\graphicspath{{figs/}{../figs/}}

\usepackage[normalem]{ulem}
\usepackage[numbers]{natbib} 
\usepackage[font=footnotesize]{caption}
\usepackage{subcaption}
\usepackage{enumitem} 
\setlist{           
  listparindent=\parindent,
  parsep=0pt,
}

\def\P{{\mathsf P}}
\def\E{{\mathsf E}}

\DeclareMathOperator*{\Bin}{\mathrm{Binom}}
\DeclareMathOperator*{\Bern}{\mathrm{Bern}}

\DeclarePairedDelimiter{\floor}{\lfloor}{\rfloor}
\newcommand{\er}{Erd\H{o}s--R\'{e}nyi}

\newcommand{\prob}{\mathsf{P}}

\newcommand{\V}{\mathcal{V}}
\newcommand{\vu}{\mathcal{V}^{\mathrm{u}}}
\newcommand{\va}{\mathcal{V}^{\mathrm{a}}}

\newcommand{\sa}{s_\mathrm{a}}
\newcommand{\su}{s_\mathrm{u}}
\newcommand{\vuone}{\mathcal{V}_1^\text{u}}
\newcommand{\vutwo}{\mathcal{V}_2^\text{u}}

\newcommand{\Na}{\mathcal{N}^\text{a}}
\newcommand{\DKL}{D_\mathrm{KL}}
\newcommand{\AttrRich}{\textsc{AttrRich}}
\newcommand{\AttrSparse}{\textsc{AttrSparse}}
\usepackage{amsthm}
\usepackage{thmtools}
\makeatletter
\newtheorem*{rep@theorem}{\rep@title}
\newcommand{\newreptheorem}[2]{%
\newenvironment{rep#1}[1]{%
 \def\rep@title{#2 \ref{##1}}%
 \begin{rep@theorem}}%
 {\end{rep@theorem}}}
\makeatother

\newtheorem{theorem}{Theorem}
\newreptheorem{theorem}{Theorem}
\newtheorem{lemma}{Lemma}
\newreptheorem{lemma}{Lemma}

\newreptheorem{fact}{Fact}
\newtheorem{corollary}{Corollary}

\theoremstyle{definition}

\newtheorem{remark}{Remark}

\hypersetup{
    colorlinks=true,
    linkcolor=black,
}

\usepackage{tikz}
\DeclareRobustCommand*\circled[1]{\tikz[baseline=(char.base)]{
    \node[shape=circle, draw, inner sep=.05pt, 
        minimum height=3pt] (char) {#1};}}
\newcommand{\indi}{\mathbbm{1}}
\title{On the Feasible Region of Efficient Algorithms for Attributed Graph Alignment}

\author{Ziao Wang, Ning Zhang, Weina Wang, and Lele Wang
\thanks{Ziao Wang is with the Department of Electrical and Computer Engineering, University of British Columbia, Vancouver, BC V6T1Z4, Canada (email: ziaow@ece.ubc.ca).}
\thanks{Ning Zhang is with the Department of Electrical and Computer Engineering, University of British Columbia, Vancouver, BC V6T1Z4, Canada (email: ningz@ece.ubc.ca).}
\thanks{Weina Wang is with the Computer Science Department, Carnegie Mellon University, Pittsburgh, PA 15213, USA, (email: weinaw@cs.cmu.edu).}
\thanks{Lele Wang is with the Department of Electrical and Computer Engineering, University of British Columbia, Vancouver, BC V6T1Z4, Canada (email: lelewang@ece.ubc.ca).}
\thanks{This work was presented in part at the 2022 IEEE International Symposium on Information Theory.}}

\allowdisplaybreaks

\begin{document}

\maketitle

\begin{abstract}
Graph  alignment  aims  at  finding  the  vertex  correspondence between two correlated graphs, a task that frequently occurs in graph mining applications such as social network analysis.
Attributed graph alignment is a variant of graph alignment, in which publicly available side information or attributes are exploited to assist graph alignment. Existing studies on attributed graph alignment focus on either theoretical performance without computational constraints or empirical performance of efficient algorithms. This motivates us to investigate efficient algorithms with theoretical performance guarantee.
In this paper, we propose two polynomial-time algorithms that exactly recover the vertex correspondence with high probability. 
The feasible region of the proposed algorithms is \emph{near optimal} compared to the information-theoretic limits. 
When specialized to the seeded graph alignment problem under the seeded \er{} graph pair model, the proposed algorithms \emph{extends} the best known feasible region for exact alignment by polynomial-time algorithms.
\end{abstract}

\section{Introduction}
The graph alignment problem, also referred to as the graph matching or noisy graph isomorphism problem, is the problem of finding the correspondence between the vertices of two correlated graphs. This problem has been given increasing attention for its applications in social network de-anonymization. 
{For instance, datasets of social networks are typically anonymized for privacy protection.
However, an attacker may be able to de-anonymize the dataset by aligning its user-user connection graph with that of publicly available data.}
{\emph{Attributed graph alignment} is a variant of graph alignment in which side information, referred to as attributes of vertices, is also publicly available in addition to the user-user connection information.}
This variant is motivated by the fact that there might exist publicly available attribute information in social networks. For example, the anonymized network Netflix has users' movie-watching history and ratings publicly available. 
Moreover, the examination of the proposed model provides both achievability and converse results for various well-known variations of the \er{} graph alignment problems. These variations include the traditional graph alignment problem without attributes, the seeded graph alignment problem, and the bipartite alignment problem. Consequently, the study of attributed graph alignment introduces a novel perspective for investigating these problems within an integrated framework.


In this paper, we focus on the attributed graph alignment problem under the attributed {\er} pair model $\mathcal{G}(n,p,\su; m,q,\sa)$, first proposed in~\cite{Zhang--Wang--Wang2021}. In this model, a base graph $G$ is generated on the vertex set $[n+m]$ where the vertices from the set $[n]$ represent \emph{users} and the rest of the vertices represent \emph{attributes}. Between each pair of users, an edge is generated independently and identically with probability $p$ to represent their connection. For each user-attribute pair, an edge is generated independently and identically with probability $q$ to represent their association. Note that there are no edges between attributes. The graph $G$ is then independently subsampled to two graphs $G_1$ and $G_2$, where each user-user edge is subsampled with probability $\su$ and each user-attribute edge is subsampled with probability $\sa$. 
To model the anonymization procedure, a random permutation $\Pi^*$ chosen uniformly at random is applied to the \emph{users} in $G_2$ to generate an anonymized version $G_2'$. Our goal in this model is to achieve exact alignment, i.e., to exactly recover the permutation $\Pi^*$ using $G_1$ and $G_2'$.

For the attributed graph alignment problem, and the graph alignment problem in general, two often asked questions are the following.
First, \emph{for what region of graph statistics is exact alignment feasible with unlimited computational power?} This region is usually referred to as the information-theoretically feasible region or the information-theoretic limits.
Second, \emph{for what region of graph statistics is exact alignment feasible with polynomial-time algorithms?}
This region is usually referred to as the feasible region of polynomial-time algorithms.
Characterizing these two feasible regions and their relationship is of utmost importance to developing a fundamental understanding of the graph alignment problem.

There has been extensive studies on these two questions under the {\er} pair model without attributes. A line of research focuses on the information-theoretic limits of exact alignment \cite{Pedarsani-Grossglauser2011,Cullina-Kiyavash2016,Cullina-Kiyavash2017,Settling-TIT}. 
Roughly speaking, it is shown that exact alignment is information-theoretically feasible when the intersection graph is dense enough. A sharp threshold of exact alignment has been established, while there still exists some gap between the converse and the achievability results. Another line of research focuses on polynomial-time algorithms for exact alignment~\cite{Dai-Cullina-Kiyavash-Grossglauser2019,Ding-Ma-Wu-Xu2020,Fan2020,Mao-Rudelson-Tikhomirov2021,chandelier2022}. Compared to the information-theoretic limits, the existing polynomial-time algorithms further require higher edge correlation between the pair of graphs to achieve exact alignment. 
The question of whether there exists polynomial-time algorithms that achieve the known information-theoretic limits is still left open.

For the attributed graph alignment problem, the information-theoretic limit has been studied in~\cite{Zhang--Wang--Wang2021}, where the feasible region (achievability results) and infeasible region (converse results) are characterized with a gap in between in some regimes.
However, the feasible region of polynomial-time algorithms for attributed graph alignment has not been studied before, and it is the focus of this paper.
In this work, we propose two polynomial-time algorithms for attributed graph alignment and characterize their feasible regions.
The two algorithms are designed for two different regimes of parameters based on the richness of attribute information:
the algorithm \AttrRich{} is designed for the regime where $mq\sa^2=\Omega(\log n)$, referred to as the attribute-information rich regime; and the algorithm \AttrSparse{} is designed for the regime where $mq\sa^2=o(\log n)$, referred to as the attribute-information sparse regime.
In both algorithms, we first explore the user-attribute connections to align a set of anchor users, and then utilize the user-user connections to the anchors to align the rest of users. 
{Due to the regime difference, \AttrRich{} is able to generate a much larger set of anchors in the first step than \AttrSparse{}.
Therefore, \AttrRich{} and \AttrSparse{} make use of the anchors differently in the second step: \AttrRich{}}
explores one-hop user-user connections to align the rest of users, while 
\AttrSparse{} explores one-hop or multiple-hop user-user connections to align the rest of users depending on the sparsity of user-user connections. This idea of matching vertices based on common neighbor witnesses has been explored to construct efficient graph alignment algorithms under the context of seeded graph alignment~\cite{Korula--Lattanzi2014,Yartseva2013,Mossel2020,Shirani2017}. In this work, we employ this idea in a two-step procedure under the setting of attributed graphs, and analyze its performance through a careful treatment of the dependency between the two steps.

\begin{figure}[htbp]
\hspace{-30pt}
\centering
    \def\svgwidth{0.5\columnwidth}
    \input{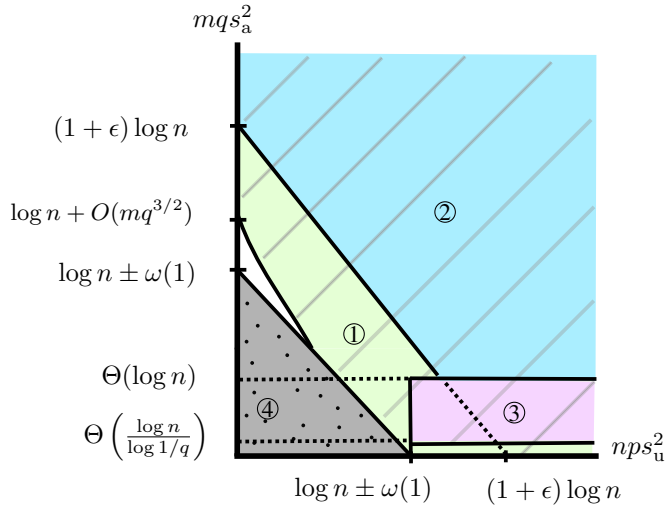}
    \caption{Comparison between the feasible regions of the proposed algorithms and the information-theoretic limits: 
    the shaded area (\circled{\small{1}}+\circled{\small{2}}+\circled{\small{3}}) represents the information-theoretically feasible region given in \cite{Zhang--Wang--Wang2021}; 
    area \circled{\small{2}} is the feasible region for Algorithm \AttrRich{} and area \circled{\small{3}} is the feasible region for Algorithm \AttrSparse{}; 
    area \circled{\small{4}} is the information-theoretically infeasible region given in \cite{Zhang--Wang--Wang2021}.}
    \label{fig:info-comparison}
\end{figure}

Our characterizations of the feasible regions of \AttrRich{} and \AttrSparse{} are illustrated in Figure~\ref{fig:info-comparison} as areas \circled{\small{2}} and \circled{\small{3}}, respectively.
The information-theoretically feasible and infeasible regions given in \cite{Zhang--Wang--Wang2021} are also illustrated in the figure for comparison.
We can see that there is a gap between the feasible region achieved by \AttrRich{} and \AttrSparse{} and the known information-theoretically feasible region.
It is left open whether this gap is a fundamental limit of polynomial-time algorithms
\footnote{We comment that in the line of work~\cite{Zhang-Tong2016,Zhang-Tong-2019,Zhou-Li-Wu-Cao-Ying-Tong2021}, the authors have explored efficient algorithms for a closely-related problem known as attributed network alignment. In this problem, attributes are attached to both vertices and edges, in contrast to the exclusive association with vertices in attributed graph alignment. The focus in this line of work is the empirical performance rather than the theoretical feasible region. } 


In addition, we specialize the feasible region of the proposed algorithms to the context of the seeded \er{} graph alignment problem, and demonstrate that the specialized feasible region includes certain range of parameters that is not known to be achievable in the literature.

\begin{figure*}[htbp]
\centering
\begin{subfigure}{0.45\textwidth}
    \def\svgwidth{1.2\columnwidth}
    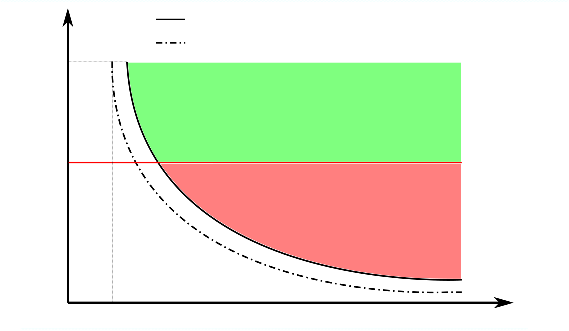
    \caption{Feasible region of polynomial-time algorithms when $mq\sa^2=0$}
    \label{fig:no_attri}
\end{subfigure}
\hfill
\begin{subfigure}{0.45\textwidth}
    \def\svgwidth{1.2\columnwidth}
    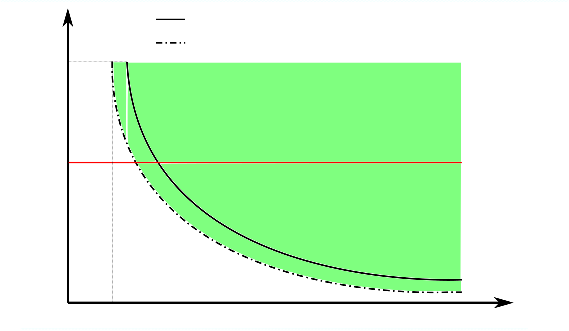
    \caption{Feasible region of polynomial-time algorithms when $mq\sa^2=0.1\log n$ }
    \label{fig:with_attri}
\end{subfigure}
\caption{Comparison between feasible regions of polynomial-time algorithms when $mq\sa^2=0$ and when $mq\sa^2=0.1\log n$. Subgraph (a) captures the case when $mq\sa^2=0$. The green region is known to be feasible by a polynomial-time algorithm in~\cite{chandelier2022}, while no polynomial-time algorithms are known to be feasible in the red region. Subgraph (b) captures the case when $mq\sa^2=0.1\log n$. The green region is feasible by the proposed algorithm \AttrRich{}.}
\label{fig:attri-effect}
\end{figure*}

Our results reveal that attributes possibly facilitate graph alignment in a much more significant way when computational efficiency is a priority.
We demonstrate this possible impact of attributes under the sparse regime $np=\Theta(\log n)$ in Figure~\ref{fig:attri-effect}.
In Figure~\ref{fig:no_attri}, we let $mq\sa^2=0$ so there is no information from the attributes, which is equivalent to the graph alignment problem without attributes;
in Figure~\ref{fig:with_attri}, we let $mq\sa^2=0.1\log n$.
We keep $p=o(1)$, and assume the value of $np\su/\log n$ can be an arbitrarily large constant but not tending to infinity in both settings for ease of comparison.
Comparing these two figures, we can see that when $mq\sa^2$ increases from $0$ to $0.1\log n$, the information-theoretic limits in terms of user-user parameters ($n,p$ and $\su$) improve a bit as expected.
However, a more fundamental improvement is in the feasible region achievable by polynomial-time algorithms.
In Figure~\ref{fig:no_attri}, the green region above $\su=\sqrt{\alpha}$ is achievable by a polynomial-time algorithm proposed in \cite{chandelier2022}, where $\alpha\approx 0.338$ is known as the Otter's constant.
The red region below $\su=\sqrt{\alpha}$ is not known to be achievable by any polynomial-time algorithms.
Moreover, in~\cite{sophie2023matching}, this red region is conjectured to be infeasible by any polynomial-time algorithm.\footnote{We note that the conjecture presented in~\cite{sophie2023matching} pertains to the sparse regime where $np=\Theta(\log n)$. In contrast, for the dense regime, it is conjectured in~\cite{mao2022testing} that no polynomial-time algorithm can achieve exact alignment if $\su\le 1/\mathrm{polylog}(n)$.}
In comparison, in Figure~\ref{fig:with_attri}, 
the green region above the information theoretic limit, which is presented as the dotted curve, is achievable by our proposed polynomial-time algorithm \AttrRich{} (assuming $q=o(1)$ and $\sa=\Theta(1)$).
Therefore, if $\sqrt{\alpha}$ is indeed the right threshold between achievable and impossible under polynomial-time algorithms for graph alignment without attributes in the sparse regime, a small amount of attribute information lowers this threshold to an arbitrarily small constant.

\section{Model}
\label{sec:model}

In this section, we describe a random process that generates a pair of correlated graphs, which we refer to as the attributed \er\ pair model $\mathcal{G}(n,p,\su; m,q,\sa)$. Under this model, we define the exact alignment problem. 

\begin{figure}[htbp]
    \centering
    \includegraphics[width=0.4\linewidth]{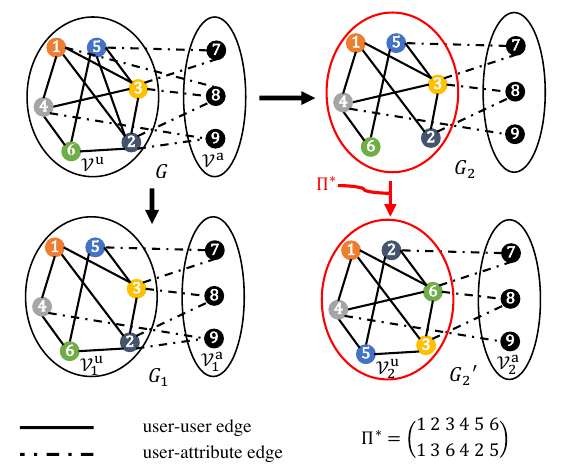}
    \caption{An illustration of  attributed \er{} pair model. We first sample a base graph $G$. Then we get  $G_1$ and $G_2$ through edge subsampling $G$. The anonymized graph $G_2'$ is obtained through apply the permutation $\Pi^*$ on the user vertex set of $G_2$. }
    \label{fig:subsampling}
\end{figure}

\emph{\textbf{Base graph generation.}} We first generate a base graph $G$, whose vertex set {$\V(G)$} consists of two disjoint sets, the \emph{user vertex set} {$\vu = \{1, 2,\ldots, n\}$} and the \emph{attribute vertex set} {$\va = \{n+1, n+2, \ldots, n+m \}$.} 
There are two types of edges in the base graph $G$, {the} user-user edges (edges connecting a pair of users) and {the} user-attribute edges (edges connecting a user vertex and an attribute vertex). The user-user edges are generated independently and identically with probability $p$, and the user-attribute edges are generated independently and identically with probability $q$. 
Throughout this paper, we assume that $p=o(1)$ and $q=o(1)$.  We write $i\stackrel{G}{\sim}j$ if vertices $i$ and $j$ are connected in graph $G$.

\emph{\textbf{Edge subsampling.}} From the base graph $G$, we obtain two correlated graphs $G_1$ and $G_2$ by subsampling the edges in $G$ independently. More specifically, we get $G_1$ and $G_2$ by independently including each user-user edge in $G$ with probability $\su$ and independently including each user-attribute edge with probability $\sa$. Throughout this paper, we assume that $\su = \Theta(1)$ and $\sa = \Theta(1)$. 

\emph{\textbf{Anonymization.}} From the $G_2$ generated as above, we get an anonymized graph $G_2'$ by applying an unknown permutation $\Pi^*$ on the user vertices of $G_2$, where $\Pi^*$ is drawn uniformly at random from the set of all possible permutations on $\vu$. 
We use $\vutwo$ to denote the user vertex set of $G_2'$ and use  $\vuone$ to denote the user vertex set of $G_1$.
Finally, we remark that this subsampling process is a special case of an earlier described attributed \er\ pair model in \cite{Zhang--Wang--Wang2021}.

\emph{\textbf{Exact alignment.}} Given an observable pair $(G_1,G_2')$, our goal is to recover {the unknown permutation $\Pi^*$, which allows us to recover the original labels of user vertices in the anonymized graph $G_2'$.} 
We say exact alignment is achieved \emph{with high probability} (w.h.p.) if $\lim_{n \rightarrow \infty}\prob(\hat{\Pi} \neq \Pi^*) = 0$.
It is worth mentioning that $\P(\hat{\Pi} \neq \Pi^*) = \P(\hat{\Pi} \neq \Pi^*|\Pi^* = \pi_\text{id})$ due to the symmetry among user vertices. Thus, we later assume without loss of generality that the {underlying} true permutation is the {identity} permutation.

\emph{\textbf{Related random graph models}.} A closely related graph generation model is the correlated stochastic block model~\cite{Onaran2016, cullina2016simultaneous, lyzinski2015graph}. In this model, the base graph $G$ is instead generated from the stochastic block model, where vertices are grouped into latent blocks, and edges are formed between blocks with specific probabilities, capturing the underlying community structure in the network. The base graph $G$ is then subsampled into the two generated graphs $G_1$ and $G_2$. In the proposed attributed \er{} graph pair models, users and attributes can be viewed as two communities in the correlated stochastic block model. However, a key distinction lies in the fact that the attributed \er{} graph pair model specifies the correspondence between attributes in the two graphs, whereas the correlated stochastic block model does not disclose such information.

Another well-studied graph alignment model with side information is the seeded \er{} graph pair model~\cite{Korula--Lattanzi2014,Yartseva2013,Mossel2020,Shirani2017}, where we have access to part of the true correspondence between user vertices.
To make a comparison between the seeded \er{} graph pair model and the proposed model, here we describe the seeded \er{} pair model $\mathcal{G} (N, \alpha, p,s)$ in detail. We first sample a base graph $G$ from the \er\ graph on $N$ vertices with edge probability $p$. Then two correlated copies $G_1$ and $G_2$ are obtained by independently subsampling the edges in the base graph where each edge is preserved with probability $s$. The anonymized graph $G_2'$ is obtained by applying an unknown permutation $\Pi^*$ on $G_2$, where $\Pi^*$ is drawn uniformly at random. Let $\mathcal{V}(G_1)$ and $\mathcal{V}(G_2')$ denote the vertex sets of $G_1$ and $G_2'$ respectively. Then, a subset $\mathcal{V}_s \subset \mathcal{V}(G_1)$ of size $\floor{N\alpha}$ is chosen uniformly at random and we define the vertex pairs $\mathcal{I}_0 = \{(v_1,\Pi^*(v_1)): v_1 \in \mathcal{V}_s\}$ as the \emph{seed set}. The graph pair $(G_1, G_2')$ together with the
seed set $\mathcal{I}_0$ are given and the goal of the exact alignment is to recover the underlying permutation for the remaining vertices w.h.p. 

Comparing the seeded \er{} pair model and the attributed \er\ pair model, we can see that the seed set and the attribute set both provide side information to assist the alignment of the remaining vertices. Nevertheless, there are two main differences between the two models. First, in the attributed \er\ pair model, we allow different edge probabilities and subsampling probabilities for user-user edges and user-attribute edges, whereas in the seeded \er\ pair model, the edge probability is identical for all edges and so is the subsampling probability.
Second, while there are edges between seeds in seeded \er{} pair model, there are no attribute-attribute edges in the attributed \er\ pair model. However, it can be shown that the existence of edges between seeds has no influence  on the information-theoretic limits for exact alignment in the seeded \er{} pair model (see Lemma 1 in~\cite{Zhang-Wang-Wang-arxiv}).
This further suggests that regarding the task of exact alignment, the information-theoretic limits on  attributed graph alignment recover the information-theoretic limits on seeded \er{} graph alignment if we specialize $p=q$ and $\su=\sa$ in the attributed \er\ pair model {$\mathcal{G}(n,p,\su;m, q,\sa)$}.

\emph{\textbf{Other notation.}}
Our algorithms rely on exploring the neighborhood similarity of user vertices in $G_1$ and $G_2'$. Here we introduce our notation of local neighborhoods. 
We define $\mathcal{N}_1^\text{a}(i)\triangleq\{j\in\mathcal{V}_1^\text{a}: i\stackrel{G_1}{\sim}j\}$ as the set of attribute neighbors of a user vertex $i$ in $G_1$ and $\mathcal{N}_2^\text{a}(i)\triangleq\{j\in\mathcal{V}_2^\text{a}: i\stackrel{G_2'}{\sim}j\}$ as the set of attribute neighbors of a user vertex $i$ in $G_2'$.
For two user vertices $i$ and $j$ in the same graph, let $d(i,j)$ be the length of the shortest path connecting $i$ and $j$ via user-user edges. For a user vertex $i\in \mathcal{V}_1^\text{u}$, we define the set of $l$-hop user neighbors of vertex $i$ as $\mathcal{N}_1^\text{u}(i,l) \triangleq \{j \in \mathcal{V}_1^{\text{u}}: d(i,j) \le l\}$ for any positive integer $l$. By convention, when $l = 1$, we simply write {$\mathcal{N}_1^\text{u}(i) \equiv \mathcal{N}_1^\text{u}(i,1)$}. The quantities $\mathcal{N}_2^\text{u}(i,l)$ and $\mathcal{N}_2^\text{u}(i)$ are defined similarly for user vertices in $G_2'$.

\emph{\textbf{Reminder of the Landau notation.}}
\begin{table}[!h]
\centering
\vspace{-0.05in}
\begin{tabular}{ c c } 
 \toprule
 Notation & Definition\\ 
 \midrule
 $f(n)= \omega(g(n))$ & $\lim\limits_{n\rightarrow\infty} \frac{|f(n)|}{g(n)} = \infty$ \\
 $f(n) = o(g(n))$ & $\lim\limits_{n \to \infty} \frac{|f(n)|}{g(n)} = 0$\\ 
 $f(n)= O(g(n))$ & $\limsup\limits_{n\rightarrow\infty}\frac{|f(n)|}{g(n)} < \infty$\\
 $f(n) = \Omega(g(n))$ & $\liminf\limits_{n\rightarrow \infty} \frac{|f(n)|}{g(n)} >0$\\
 $f(n) = \Theta(g(n))$  & $f(n) = O(g(n))$ and $f(n) = \Omega(g(n))$\\
 \bottomrule
\end{tabular}
\end{table}

\vspace{2pt}

\section{Main results}

In this section, we propose two polynomial-time algorithms for the attributed graph alignment problem. Their feasible regions are characterized in the following two theorems.

\begin{theorem}\label{thm:achievability}
Consider the attributed \er{} pair $\mathcal{G}(n,p,s_\mathrm{u}; m, q,s_\mathrm{a})$ with $p=o(1)$, $q=o(1)$, $s_\mathrm{u}=\Theta(1)$, and $s_\mathrm{a}=\Theta(1)$. Assume that
\begin{equation}
    mqs_\mathrm{a}^2=\Omega(\log n)
    \label{eq:cond1}
\end{equation}
and that there exists some constant $\epsilon>0$ such that
\begin{equation}
    mqs_\mathrm{a}^2+nps_\mathrm{u}^2\ge (1+\epsilon)\log n.
    \label{eq:cond2}
\end{equation}
Then there exists a polynomial-time algorithm, namely, Algorithm~\AttrRich{} with the parameters chosen in~\eqref{eq:x} and \eqref{eq:y}, that achieves exact alignment w.h.p.  
\end{theorem}
\begin{theorem}\label{thm:achievability2}
Consider the attributed \er{} pair $\mathcal{G}(n,p,s_\mathrm{u}; m, q,s_\mathrm{a})$ with $p=o(1)$, $q=o(1)$, $\su=\Theta(1)$, and $\sa=\Theta(1)$. Assume that
\begin{equation}
    mq\sa^2=o(\log n),
    \label{eq:ac2_cond1}
\end{equation}
\begin{equation}
    np\su^2-\log n\rightarrow +\infty,
    \label{eq:ac2_cond2}
\end{equation}
and that there exists some constant $\tau>0$ such that
\begin{equation}
    mq\sa^2\ge\frac{2\log n}{\tau\log \frac1q}.
    \label{eq:ac2_cond4}
\end{equation}
Then there exists a polynomial-time algorithm, namely, Algorithm~\AttrSparse{} with the parameters chosen in~\eqref{eq:algo2-para1},~\eqref{eq:algo2-para2} and~\eqref{eq:algo2-para3}, that achieves exact alignment w.h.p.  
\end{theorem}
The proofs of Theorems~\ref{thm:achievability}~and~\ref{thm:achievability2} are deferred to Sections~\ref{sec:proof-thm1}~and~\ref{sec:proof-thm2} respectively.

\subsection{Algorithm \AttrRich{}}
\label{alg-1}
In this subsection, we propose the first algorithm that leads to the achievable region in Theorem~\ref{thm:achievability}. This algorithm is designed for the attribute-information rich regime, hence named \AttrRich{}.
\begin{itemize}
    \item {\bf Input:} The graph pair $(G_1, G_2')$ and thresholds $x$ and $y$.
    \item {\bf Step 1: Align through attribute neighbors.} In this step, we only consider the edge connections between users and attributes, and use these information to find the matching for a set of vertices which will be later referred to as anchors. For each pair of users $i\in \vuone$ and $j \in \vutwo$, compute the number of common attribute neighbors
\begin{equation}
C_{ij} \triangleq |\Na_1(i) \cap \Na_2(j)|.
\label{eq:cij}
\end{equation}
If $C_{ij} > x$, add $(i,j)$ into $\mathcal{S}_1$. We refer to vertex pairs in the set $\mathcal{S}_1$ as anchor. If there exists conflicting pairs in $\mathcal{S}_1$, i.e., two distinct pairs $(i_1,j_1)$ and $(i_2,j_2)$ with $i_1 = i_2$ or $j_1 = j_2$, set $\mathcal{S}_1 = \emptyset$ and declare failure. Otherwise, set $\hat{\pi}(i) = j$ for all pairs $(i,j) \in \mathcal{S}_1$.
\item {\bf Step 2: Align through user neighbors.} In the previous step, we have aligned the anchors using the edges between users and attributes. In this step, we will align the non-anchor vertices by their edge connections to the anchors. Let 
\[\mathcal{U}_1\triangleq\{i\in\mathcal{V}_1^\text{u}:(i,j)\not\in \mathcal{S}_1,\forall j\in\mathcal{V}_2^\text{u}\}
\]
denote the set of all unmatched vertices in $G_1$ and let 
\[
\mathcal{U}_2\triangleq\{j\in\mathcal{V}_2^\text{u}:(i,j)\not\in \mathcal{S}_1,\forall i\in\mathcal{V}_1^\text{u}\}
\]
denote the set of all unmatched vertices in $G_2$. 
For each unmatched pair $i \in \mathcal{U}_1$ and $j \in \mathcal{U}_2$, consider the user neighbors of $i$ and the user neighbors of $j$ that are matched as pairs in $\mathcal{S}_1$, and compute the number of such matched pairs for $(i,j)$
\begin{equation}
W_{ij}\triangleq\sum_{k\in \mathcal{N}_1^\text{u}(i), l\in \mathcal{N}_2^\text{u}(j)}\indi_{\{(k,l)\in \mathcal{S}_1\}}.
\label{eq:wij}
\end{equation}
For each $i \in \mathcal{U}_1$, if $W_{ij} > y|\mathcal{S}_1|$ for a unique $j \in \mathcal{U}_2$, 
set $\hat{\pi}(i)=j$. Otherwise, declare failure.
If $\hat{\pi}$ is not a bijection from $\vuone$ to $\vutwo$, declare failure.
\item {\bf Output:} The estimated permutation $\hat{\pi}$.    
\end{itemize}

In this algorithm, there are two threshold parameters $x$ and $y$. 
In the following analysis, we choose 
\begin{equation}
x=(1-\delta_x)mqs_\text{a}^2,
\label{eq:x}
\end{equation}
where $1-\delta_x=\frac{\Delta_x}{\log\frac{1}{q}}$ with constant $\Delta_x\ge\max\{1,\frac{3\log n}{mqs_\text{a}^2}\}$, and 
\begin{equation}
y=(1-\delta_y)ps_\text{u}^2,
\label{eq:y}
\end{equation}
where $1-\delta_y=\frac{\Delta_y}{\log\frac{1}{p}}$ with constant $\Delta_y\ge 2$.

\begin{remark}[Complexity of Algorithm \AttrRich{}]
In Algorithm~\AttrRich{}, the time complexity for computing $C_{ij}$ for all pairs $(i,j)\in \vuone\times\vutwo$ is $O(n^2m)$ since there are $n^2$ pairs and for each pair, there are $m$ attributes to consider. Similarly, the time complexity for computing $W_{ij}$ for all pairs $(i,j)\in \mathcal{U}_1\times\mathcal{U}_2$ is at most $O(n^3)$. Therefore, if $m=\omega(n)$, the time complexity of Algorithm~\AttrRich{} is $O(n^2m)$ and if $m=O(n)$, the time complexity of Algorithm~\AttrRich{} is $O(n^3)$. 
\end{remark}

\subsection{Algorithm \AttrSparse{}} 
\label{alg-2}
In this subsection, we propose the second algorithm that leads to the achievable region in Theorem~\ref{thm:achievability2}. This algorithm is designed for the attribute-information sparse regime, hence named \AttrSparse{}.
In Step 2 of this algorithm, we consider two different cases. In the case when the user-user connection is dense, we perform a similar process as in Step 2 of Algorithm~\AttrRich{}. In the case when the user-user connections is sparse, we call a seeded alignment algorithm proposed in~\cite{Mossel2020}, which is restated in Subsection~\ref{alg:Jiaming3}.
\begin{itemize}
    \item {\bf Input:} The graph pair $(G_1,G_2')$, three thresholds $y$, $z$ and $\eta$, an integer $l$, and the model parameters $n$ and $p$.
    \item {\bf Step 1: Align through attribute neighbors.} Similar to Step 1 of Algorithm~\AttrRich{}, for each pair of users $i\in \vuone$ and $j \in \vutwo$, we compute the quantity
\begin{equation}
C_{ij} = |\Na_1(i) \cap \Na_2(j)|.
\end{equation}
Unlike Step 1 of Algorithm~\AttrRich{}, we create an anchor set using a different threshold $z$.
If $C_{ij} > z$, add $(i,j)$ into $\mathcal{S}_2$. We refer to vertex pairs in the set $\mathcal{S}_2$ as anchors. If there exists conflicting pairs in $\mathcal{S}_2$, i.e., two distinct pairs $(i_1,j_1)$ and $(i_2,j_2)$ with $i_1 = i_2$ or $j_1 = j_2$, set $\mathcal{S}_2 = \emptyset$ and declare failure.
    \item {\bf Step 2: Align through user-user edges.} 
    \begin{itemize}
    
    \item If $np> n^{1/7}$, we perform the similar process as in Step 2 of Algorithm \AttrRich{} to align the non-anchor vertices. Define
    \[\mathcal{U}_3\triangleq\{i\in\mathcal{V}_1^\text{u}:(i,j)\not\in \mathcal{S}_2,\forall j\in\mathcal{V}_2^\text{u}\}
    \]  
    and
    \[
    \mathcal{U}_4\triangleq\{j\in\mathcal{V}_2^\text{u}:(i,j)\not\in \mathcal{S}_2,\forall i\in\mathcal{V}_1^\text{u}\}.
    \]
    For each unmatched pair $i\in\mathcal{U}_3$ and $j\in\mathcal{U}_4$, compute $W_{ij}$ as defined in~\eqref{eq:wij}. For each $i\in\mathcal{U}_3$, if $W_{ij}>y|\mathcal{S}_2|$ for a unique $j\in\mathcal{U}_4$, set $\hat{\pi}(i)=j$. Otherwise, declare failure. If $\hat{\pi}$ is not a bijection from $\vuone$ to $\vutwo$, declare failure.

    \item If $np \le n^{1/7}$, run Algorithm~\ref{alg:Jiaming3} with the induced subgraphs on user vertices $\mathcal{V}_1^\text{u}$ and $\mathcal{V}_2^\text{u}$, the seed set $\mathcal{I}_0=\mathcal{S}_2$, and parameters $l$ and $\eta$. 

    \end{itemize}

   \item {\bf Output:} The estimated permutation $\hat{\pi}$.
    
    
\end{itemize}
In this algorithm, there are four parameters $y$, $z$, $l$ and $\eta$ that we can choose. 
In the following analysis, we choose $y$ to be the same value as in~\eqref{eq:y},
\begin{equation}
    z=(1+\tau)mq\sa^2
    \label{eq:algo2-para1}
\end{equation}
(cf. the same $\tau$ as in Theorem~\ref{thm:achievability2}),
\begin{equation}
    l=\left\lfloor\frac{(6/7)\log n}{\log(np)}\right\rfloor,
    \label{eq:algo2-para2}
\end{equation}
and
\begin{equation}
    \eta=4^{2l+2}n^{-2/7}.
    \label{eq:algo2-para3}
\end{equation}
\begin{remark}[Complexity of Algorithm \AttrSparse{}]
In Algorithm~\AttrSparse{}, the time complexity for computing $C_{ij}$ for all pairs $(i,j)\in \vuone\times\vutwo$ is $O(n^2m)$ since there are $n^2$ pairs and for each pair, there are $m$ attributes to consider. The time complexity of computing $W_{ij}$ for all pairs $(i,j)\in\mathcal{U}_3\times\mathcal{U}_4$ is $O(n^3)$ and that of Algorithm~\ref{alg:Jiaming3} is $O(n^{37/7})$ as given in~\cite{Mossel2020}, which can be improved with better data structures. Therefore, if $np>n^{1/7}$, the complexity of Algorithm~\AttrSparse{}{} is $O(n^2m+n^3)$; Otherwise its complexity is $O(n^2m+n^{37/7})$.
\end{remark}

\subsection{Seeded alignment in the sparse regime~\cite[Algorithm 3]{Mossel2020}}\label{alg:Jiaming3}
Except for the two graphs $G_1$ and $G_2'$, this algorithm takes a seed set $\mathcal{I}_0$ as input. The seed set $\mathcal{I}_0$ consists of vertex pairs $(i,j)$ such that $\pi^*(i)=j$. The algorithm utilizes this seed set to align the rest of vertices.
\begin{itemize}
    \item {\bf Input:} The graph pair $(G_1,G_2')$, the seed set $\mathcal{I}_0$, a threshold $\eta$, and an integer $l$. 
    \item {\bf Align high-degree vertices.} Let 
    \[
    \mathcal{J}_1\triangleq\{i\in\mathcal{V}(G_1):(i,j)\notin\mathcal{I}_0,\forall j\in\mathcal{V}(G_2')\},
    \]
    and
    \[
    \mathcal{J}_2\triangleq\{j\in\mathcal{V}(G_2'):(i,j)\notin\mathcal{I}_0,\forall i\in\mathcal{V}(G_1)\}.
    \]
    For each pair of unseeded vertices $u\in\mathcal{J}_1$ and $v\in\mathcal{J}_2$, and for each pair of their neighbors $i\in \mathcal{N}_1^\mathrm{u}(u)\setminus\{u\}$ and $j\in \mathcal{N}_2^\mathrm{u}(v)\setminus\{v\}$, compute
    \begin{align*}
    &\lambda_{i,j}^{u,v} =\min_{x\in\mathcal{V}(G_1),y\in\mathcal{V}(G_2')}\{|\{(k_1,k_2)\in\mathcal{I}_0:\\
    &\;k_1\in\mathcal{N}_{G_1\setminus\{u,x\}}^\mathrm{u}(i,l),k_2\in\mathcal{N}_{G_2'\setminus\{v,y\}}^\mathrm{u}(j,l)\}|\},
    \end{align*}
    where $\mathcal{N}_{G\setminus S}^\mathrm{u}(i_1,l)$ denotes the set of user vertices $i_2$ such that $d(i_1,i_2)\le l$ in the induced subgraph $G$ with the set of vertices $S$ removed. 
    Let
    \[
    Z_{u,v}=\sum_{i\in \mathcal{N}_1^\mathrm{u}(u)\setminus\{u\}}\sum_{j\in \mathcal{N}_2^\mathrm{u}(v)\setminus\{v\}}\indi_{\{\lambda_{i,j}^{u,v}\ge \eta|\mathcal{I}_0|\}}.
    \]
    If $Z_{u,v}\ge \log n /\log\log n-1$, add $(u,v)$ into set $\mathcal{T}$. Add all the vertex pairs from $\mathcal{I}_0$ to $\mathcal{T}$. If there exists conflicting pairs in $\mathcal{T}$, i.e., two distinct pairs $(i_1,j_1)$ and $(i_2,j_2)$ with $i_1 = i_2$ or $j_1 = j_2$, set $\mathcal{T} = \emptyset$ and declare failure.
    \item {\bf Align low-degree vertices.} Let 
    \[
    \mathcal{J}_3\triangleq\{i\in\mathcal{V}(G_1):(i,j)\notin\mathcal{T},\forall j\in\mathcal{V}(G_2')\},
    \]
    and
    \[
    \mathcal{J}_4\triangleq\{j\in\mathcal{V}(G_2'):(i,j)\notin\mathcal{T},\forall i\in\mathcal{V}(G_1)\}.
    \]
    For all pairs of unmatched vertices $i_1\in\mathcal{J}_3$ and $i_2\in\mathcal{J}_4$, if $i_1$ is adjacent to a user vertex $j_1$ in $G_1$ and $i_2$ is adjacent to a user vertex $j_2$ in $G_2'$ such that $(j_1,j_2)\in\mathcal{T}$, then set $\hat{\pi}(i_1)=i_2$.
    \item {\bf Finalize and output:} For each vertex pair $(i,j)\in\mathcal{T}$, set $\hat{\pi}(i)=j$. If $\hat{\pi}$ is a bijection from $\mathcal{V}(G_1)$ to $\mathcal{V}(G_2')$, output $\hat{\pi}$, otherwise declare failure.
\end{itemize}

\section{Discussion}
\label{sec:discussion}

In this section, we compare the feasible region in Theorems~\ref{thm:achievability}~and~\ref{thm:achievability2} to existing works.  In Section~\ref{sec:comparison-info}, we compare the feasible region with the information-theoretic limits in~\cite{Zhang--Wang--Wang2021}. It is shown that there still exists a gap between the feasible region of the proposed algorithms and the information-theoretic limit.
In Section~\ref{sec:comparison-seeded}, we specialize the feasible region of the proposed algorithm to the context of seeded graph alignment problem, and compare the specialized feasible region to the information-theoretic limits as well as the best-known feasible regions by polynomial-time algorithms for exact alignment in literature given in~\cite{Mossel2020}~and~\cite{Shirani2017}.
It is shown that while having a gap to the information-theoretic limit, the proposed Algorithms~\AttrRich{} and~\AttrSparse{} achieves exact recovery in certain range of parameters that is unknown to be feasible by any existing efficient algorithms.
In Section~\ref{sec:comparison-bipartite}, we consider the bipartite alignment problem which is another special case. We show that the proposed Algorithm~\AttrRich{} provides an alternative polynomial-time algorithm to the theoretically optimal Hungarian Algorithm~\cite{Kuhn1955}, with a slightly lower complexity.

\subsection{Comparison to the information theoretic limits}
\label{sec:comparison-info}
The information-theoretic limits of attributed graph alignment were established in \cite{Zhang--Wang--Wang2021}. {The version for the subsampling model is stated as follows.}
\begin{theorem}[Theorem~1 in \cite{Zhang--Wang--Wang2021}]
\label{thm:IT limits}
    Consider the attributed graph pair $\mathcal{G}(n,p,s_\mathrm{u}; m, q,s_\mathrm{a})$ with $1-p=\Theta(1)$, $1-q=\Theta(1)$, $s_\mathrm{u}=\Theta(1)$, and $s_\mathrm{a}=\Theta(1)$. \\
   \textbf{ Achievability: }
    In the regime where $q=O\left(\frac{1}{(\log n)^2}\right)$, if
    \begin{align*}
        np\su^2+mq\sa^2 \geq \log n +\omega(1), 
    \end{align*}
    then exact alignment is achievable w.h.p.\\
    In the regime where $q=\omega\left(\frac{1}{(\log n)^2}\right)$, if 
    \begin{align*}
        np\su^2+mq\sa^2 - a_n \geq \log n +\omega(1),
    \end{align*}
    where   
    $a_n \triangleq m\left(\sqrt{q\sa^2(1-q+q(1-\sa)^2)}-q\sa(1-\sa)\right)^2-mq\sa^2
    =O(mq^{3/2}),$
then exact alignment is achievable w.h.p.\\
    \textbf{Converse:} If
    \begin{align*}
        np\su^2+mq\sa^2 \leq \log n -\omega(1),
    \end{align*}
    then no algorithm achieves exact alignment w.h.p.
\end{theorem}
From Theorem~\ref{thm:IT limits} we can see that when $q=\omega\left(\frac{1}{(\log n)^2}\right)$, the achievability and converse differ by at most some constant times $mq^{3/2}$; when $q=O\left(\frac{1}{(\log n)^2}\right)$, the achievability and converse are tight, because in this regime we have $mq^{3/2}=\omega(1)$.
We visualize the information-theoretic limits and the computation feasible regions in Fig.~\ref{fig:info-comparison}.

\subsection{Specialization to the seeded \er{} graph alignment}
\label{sec:comparison-seeded} 
    In this subsection, we specialize the feasible region of proposed algorithm to the seeded \er{} graph alignment problem, and compare the feasible region with that of the existing algorithms. Consider an attributed {\er}  graph pair $(G_1,G_2')\sim \mathcal{G}(n,p,s_\text{u}; m, q,s_\text{a})$ with $p=q$ and $\sa=\su\triangleq s$. Then these $m$ attributes can be viewed as $m$ seeds and $(G_1,G_2')$ can be viewed as a graph pair generated from a seeded {\er} pair model $\mathcal{G}(n+m,\frac{m}{m+n},p,s)$, and the edges between the these $m$ vertices are all removed.\footnote{The information theoretic limit of exact alignment in the seeded graph alignment problem remains the same after removing the edges between attributes (see Lemma 1 in~\cite{Zhang-Wang-Wang-arxiv}).} For simplicity, we write 
\[
N \triangleq m+n \quad \text{and} \quad \alpha \triangleq \tfrac{m}{m+n}
\]
later on. In our comparisons, we always assume that $(1-\alpha)N=\omega(1)$, $s=\Theta(1)$ and $p=o(1)$.

We first specialize the feasible regions of Algorithm \AttrRich{} in Theorem~\ref{thm:achievability} and Algorithm \AttrSparse{} in Theorem~\ref{thm:achievability2} to the seeded graph alignment problem. 

\begin{corollary}[Feasible region of Algorithm \AttrRich{}]\label{cor:alg1}
Consider the seeded graph pair $\mathcal{G}(N,\alpha,p,s)$ with $p=o(1)$, $s=\Theta(1)$ and $(1-\alpha) N=\omega(1)$. Assume that
\begin{equation}
    \alpha Nps^2=\Omega(\log((1-\alpha)N))
    \label{eq:cor1_cond1}
\end{equation}
and that there exists some constant $\epsilon>0$ such that
\begin{equation}
    Nps^2\ge (1+\epsilon)\log ((1-\alpha)N).
    \label{eq:cor1_cond2}
\end{equation}
Then there exists a polynomial time algorithm, namely, Algorithm~\AttrRich{} with parameters
\[
x=(1-\delta_x)\alpha Nps^2 \quad\text{and}\quad y=(1-\delta_y)ps^2
\]
that achieves exact alignment w.h.p. Here, $1-\delta_x=\frac{\Delta_x}{\log\frac{1}{p}}$ with constant $\Delta_x\ge\max\{1,\frac{3\log (1-\alpha)N}{\alpha Nps^2}\}$ and $1-\delta_y=\frac{\Delta_y}{\log\frac{1}{p}}$ with constant $\Delta_y\ge 2$.

\end{corollary}
\begin{corollary}[Feasible region of Algorithm \AttrSparse{}]\label{cor:alg2}
Consider the seeded graph pair $\mathcal{G}(N,\alpha,p,s)$ with $p=o(1)$, $s=\Theta(1)$ and $(1-\alpha) N=\omega(1)$. Assume that
\begin{equation}
    \alpha Nps^2=o(\log( (1-\alpha) N)),
    \label{eq:cor2_cond1}
\end{equation}
\begin{equation}
    (1-\alpha)Nps^2-\log( (1-\alpha)N)\ge\omega(1),
    \label{eq:cor2_cond2}
\end{equation}
and that there exists some constant $\tau>0$ such that
\begin{equation}
    \alpha Nps^2\ge\frac{2\log ((1-\alpha) N)}{\tau\log \frac1p}.
    \label{eq:cor2_cond4}
\end{equation}
Then there exists a polynomial time algorithm, namely, Algorithm~\AttrSparse{} with parameters 
\begin{equation*}
    z=(1+\tau)\alpha Nps^2,
\end{equation*}
\begin{equation*}
    L=\left\lfloor\frac{(6/7)\log(1-\alpha)N}{\log(1-\alpha)Np}\right\rfloor,
\end{equation*}
and
\begin{equation*}
    \eta=4^{2l+2}((1-\alpha)N)^{-2/7},
\end{equation*}
 that achieves exact alignment w.h.p.
\end{corollary}
\begin{remark}
\label{rmk:compare_it}
As expected, the feasible region of proposed algorithms \AttrRich{} and \AttrSparse{} is a strict subset of the information-theoretic feasible region established in~\cite{Zhang--Wang--Wang2021}. We postpone the detailed comparison to Section~\ref{subsec:compare_IT}.
\end{remark}

Now we compare the proposed algorithms to the best known polynomial-time algorithms in the literature~\cite{Mossel2020} and~\cite{Shirani2017}. The comparison of their feasible regions is summarized in the following theorem, the proof of which is postponed to Section~\ref{subsec:compare_efficient}. 

\begin{theorem}[Comparison of polynomial-time algorithms]\label{thm:compare}
Consider the seeded \er{} pair model $\mathcal{G}(N,\alpha,p,s)$ with $p=o(1)$, $s=\Theta(1)$ and $(1-\alpha) N=\omega(1)$. Assume that parameters $N,\alpha,p$, and $s$ satisfy any of the following four sets of conditions:
\begin{equation*}
 \circled{1}\begin{cases}
        p\log N \log\left(\tfrac1p\right) =\omega(1),\\
        \alpha =\Omega\left(\frac{\log((1-\alpha)N)}{Nps^2\log\frac1p}\right),\\
        \alpha<\frac{2\log N}{NI(p,s)};
    \end{cases}
\end{equation*}
or
\begin{equation*}
    \circled{2}\begin{cases}
        p\log N \log\left(\tfrac1p\right) =O(1),\\
        p\log N \log^2\left(\tfrac1p\right) =\omega(1),\\
        \alpha =\Omega\left(\frac{\log((1-\alpha)N)}{Nps^2\log\frac1p}\right),\\
        \alpha =O\left(\frac{1}{NI^2(p,s)}\right);
    \end{cases}
    \end{equation*}
or
\begin{equation*}
    \circled{3}\begin{cases}
        p\log N \log^2\left(\tfrac1p\right) =O(1),\\
        Np >\frac{sN^{1/2}}{16(2-s)^2},\\
        \alpha =\Omega\left(\frac{\log((1-\alpha)N)}{Nps^2\log\frac1p}\right),\\
        \alpha <\frac{300\log N}{Nps^2};
    \end{cases}
    \end{equation*}
or 
\begin{equation*}
    \circled{4}\begin{cases}
        N =\Omega(((1-\alpha)N)^{1+\epsilon}),\\
        Nps^2-\log N =O(1),\\
        Nps^2 \ge(1+\epsilon)\log((1-\alpha)N)
    \end{cases}
    \end{equation*}
for some positive constant $\epsilon$.
Then the proposed algorithms achieve exact alignment w.h.p., while 
none the of existing algorithms  in~\cite{Mossel2020,Shirani2017} is known to achieve exact alignment w.h.p.
On the other hand, when $$\frac{\log N+\omega(1)}{s^2}\le Np\le\frac{sN^{1/2}}{16(2-s)^2},$$ the feasible region of the proposed algorithms is a strict subset of that in~\cite{Mossel2020}. 
\end{theorem}
\begin{figure}[htbp]
    \hspace{-1em}
    \centering
    \def\svgscale{1}
    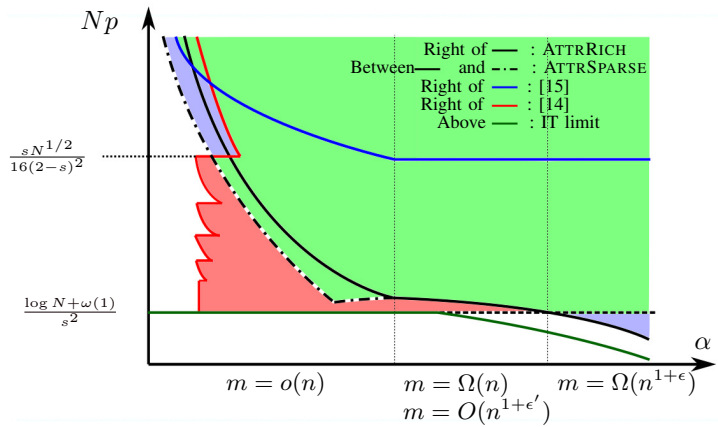
    \caption{ Comparison of the feasible region of Corollaries~\ref{cor:alg1} and~\ref{cor:alg2} to the feasible region in~\cite{Mossel2020}~and~\cite{Shirani2017}. On the top-left corner and bottom-right corner, the two blue regions are feasible for the proposed algorithms but not for any existing works. The red region is feasible for existing works, but not for the proposed algorithms. The green region is the overlap of our feasible region with the feasible region in the existing works. } 
    \label{fig:seeded-comparison}
\end{figure}
\begin{remark}
    The polynomial-time algorithms for graph alignment under the unseeded \er{} graph pair model proposed in~\cite{Mao-Rudelson-Tikhomirov2021} and~\cite{chandelier2022} trivially imply polynomial-time algorithm for seeded graph alignment. However, the feasible regions of algorithms in~\cite{Mao-Rudelson-Tikhomirov2021} and~\cite{chandelier2022} both require additional conditions on the subsampling probability $s$. Therefore, we do not include them to the comparison in this section.
\end{remark}

\begin{remark}
As shown in Corollaries~\ref{cor:alg1} and~\ref{cor:alg2}, the two proposed algorithms achieve exact alignment in two mutually exclusive regimes. Theorem~\ref{thm:compare} summarizes the comparison between the \emph{union} of the feasible regions of the two proposed algorithms and the \emph{union} of the feasible regions in~\cite{Mossel2020} and~\cite{Shirani2017}. Regions \circled{1}, \circled{2}, and \circled{3} in Theorem~\ref{thm:compare} correspond to the top left blue area in Fig.~\ref{fig:seeded-comparison}. Region \circled{4} in Theorem~\ref{thm:compare} corresponds to the bottom right blue area in Fig.~\ref{fig:seeded-comparison}.
\end{remark}
As stated in Theorem~\ref{thm:compare}, Corollaries~\ref{cor:alg1} and~\ref{cor:alg2} introduce some new feasible region in the regime $Np>\frac{s}{16(2-s)^2}N^{1/2}$ (the top-left blue corner in Fig.~\ref{fig:seeded-comparison}). 
To understand the improvement over the feasible region in~\cite{Mossel2020}, recall that the proposed algorithms in this work align vertices two steps: initially aligning a group of anchor vertices by exploring the seeds within their one-hop neighborhood, and subsequently aligning the remaining unmatched vertices with the assistance of the anchors in their respective one-hop neighborhood. 
In contrast, the algorithm proposed in \cite{Mossel2020} for this scenario aligns all vertices by exploring the one-hop seed neighbors of the non-seed vertices, which closely resembles the first step of the proposed algorithms. This is why the proposed algorithms expand the feasible region introduced in \cite{Mossel2020}.
To grasp the improvements made over the feasible region in~\cite{Shirani2017}, note that the algorithm introduced in~\cite{Shirani2017} performs a similar two-step process as the algorithms in this work. The primary distinction lies in the second step of matching. In~\cite{Shirani2017}, the matched vertices from the first step, along with the seeds, act as anchors in the second step. In contrast, our proposed algorithms utilize only the matched vertices from the first step as anchors. We comment that the improvement of the proposed algorithms over the algorithm in \cite{Shirani2017} mainly comes from the tightness of analysis.

    For completeness of the comparison, we restate the feasible regions of algorithms in~\cite{Mossel2020, Shirani2017} as follows. 
\begin{theorem}[Theorem 4 in~\cite{Mossel2020}]\label{cor:jiamingthm4}
Consider the seeded {\er} pair model $\mathcal{G}(N,\alpha,p,s)$. Suppose $Np$ can be written as $Np=bN^a$ for some constants $a$ and $b$ such that
    \[
    0<a\le 1 \quad \text{and}\quad 0<b\le\tfrac{s}{16(2-s)^2}.
    \]
Assume that
\begin{equation}
    \alpha\ge\frac{300\log N}{(Nps^2)^{\floor{1/a}}}.
    \label{eq:jiamingthm4_cond}
\end{equation}
Then there exists a polynomial-time algorithm, namely, Algorithm 2 in~\cite{Mossel2020}, that achieves exact alignment w.h.p. Moreover, the algorithm runs in $O(N^3)$ time.
\end{theorem}

\begin{theorem}[Theorem 3 in~\cite{Mossel2020}]\label{cor:jiamingthm3}
Consider the seeded {\er} pair model $\mathcal{G}(N,\alpha,p,s)$ with $Np\le N^{\beta}$ for a fixed constant $\beta<1/6$, and $s=\Theta(1)$. Assume that
\begin{equation}
    Nps^2\ge \log N +\omega(1)
    \label{eq:jiamingthm3_cond1}
\end{equation}
and
\begin{equation}
    \alpha\ge N^{-1+3\beta}.
    \label{eq:jiamingthm3_cond2}
\end{equation}
Then Algorithm~\ref{alg:Jiaming3} with the parameters 
\[
    l=\left\lfloor\frac{(6/7)\log N}{\log(Np)}\right\rfloor \quad \text{and} \quad \eta=4^{2l+2}N^{-2/7}
\]
achieves exact alignment w.h.p. Moreover, the algorithm runs in $O(n^{5+2\beta})$ time.
\end{theorem}
\begin{theorem}[Theorem 2 in~\cite{Shirani2017}]\label{thm:erkip_thm2}
Consider the seeded {\er} pair model $\mathcal{G}(N,\alpha,p,s)$ with $p=o(1)$ and $s=\Theta(1)$. Assume that 
\begin{equation}
    \alpha=\omega\left(\frac{1}{NI(p,s)^2}\right)
    \label{eq:erkip_cond1}
\end{equation}
and
\begin{equation}
    \alpha\ge \frac{2\log N}{NI(p,s)},
    \label{eq:erkip_cond2}
\end{equation}
where $I(p,s)\triangleq 2ps\log\tfrac{1}{ps}+(2-2ps)\log \tfrac{1}{1-ps}+ps^2\log ps^2+2ps(1-s)\log (ps-ps^2)+(1+ps^2-2ps)\log(1+ps^2-2ps)  =(1+o(1))s^2p\log\tfrac1p $ denotes the mutual information between a pair of correlated edges in $G_1$ and $G_2'$. Then there exists a polynomial-time algorithm, namely, the TMS algorithm in~\cite{Shirani2017}, that achieves exact alignment w.h.p.
\end{theorem}

\subsection{Specialization to the bipartite alignment}
\label{sec:comparison-bipartite}
Consider an attributed {\er} graph pair $(G_1,G_2)\sim \mathcal{G}(n,p,s_\text{u}; m, q,s_\text{a})$ with $p=0$ or $\su=0$. Then $G_1$ and $G_2$ reduce to two bipartite graphs with edges connected only between users and attributes. In this special case, conditions~\eqref{eq:cond1} and~\eqref{eq:cond2} reduce to a single condition: If there exists some positive constant $\epsilon>0$ such that
\begin{equation}
    mq\sa^2\ge(1+\epsilon)\log n,
    \label{eq:cond_bipartite}
\end{equation}
then Algorithm~\AttrRich{} achieves exact alignment w.h.p.
In contrast, Corollary 1 in~\cite{Zhang--Wang--Wang2021} implies that the maximum likelihood estimator exactly recovers $\pi^*$ w.h.p if 
\begin{equation}
    mq\sa^2\ge\log n+\omega(1).
    \label{eq:cond_hungarian}
\end{equation}
Moreover, the maximum likelihood estimator can be computed in polynomial time by first computing the similarity score for each pair of vertices $(u,v)\in \vuone\times\vutwo$ as
$Z_{u,v} = |\mathcal{N}_1^\text{a}(u)\cap\mathcal{N}_2^\text{a}(v)|$
and then solving the balanced assignment problem using the famous Hungarian Algorithm first proposed in~\cite{Kuhn1955}. From conditions~\eqref{eq:cond_bipartite} and~\eqref{eq:cond_hungarian}, we can see that the feasible region of Algorithm~\AttrRich{} is completely covered by the feasible region of the Hungarian Algorithm. By the above argument, the maximum likelihood estimator requires $O(n^2m)$ time complexity to compute the similarity score for all pairs of vertices and the Hungarian Algorithm can be implemented with $O(n^3)$ time complexity~\cite{Munkres1957}. Therefore, if $m=\omega(n)$, the time complexity of the maximum likelihood estimator is $O(n^2m)$ and if $m=O(n)$, the time complexity of the maximum likelihood estimator is $O(n^3)$. In comparison, the time complexity of Algorithm~\AttrRich{} in this special case is always $O(n^2m)$.
\section{Proof of Theorem~\ref{thm:achievability}}\label{sec:proof-thm1}
Recall that our algorithm consists of the following two steps: Step~1 (Align through attribute neighbors) produces a set $\mathcal{S}_1$ that we refer to as the set of anchor pairs based on the $C_{ij}$ defined in \eqref{eq:cij};  Step~2 (Align through user neighbors) aligns the remaining vertices based on the $W_{ij}$ defined in \eqref{eq:wij}. Moreover, recall that we assume without generality that the true underlying permutation $\pi^*$ is the identity permutation.
Our proof of Theorem~\ref{thm:achievability} analyzes the following corresponding error events.

\begin{itemize}[leftmargin=1em]
\item \textbf{Step 1 Error Event.}
Define
\[
\mathcal{E}_1\triangleq\{\exists (i,j)\in\mathcal{V}_1^\text{u}\times\mathcal{V}_2^\text{u}\text{ s.t. }i\neq j\text{ and }C_{ij}>x\}.
\]
The event $\mathcal{E}_1^c$ guarantees that the anchor set found in the first step only contains correctly matched pairs.

\item \textbf{Step~2 Error Events.}
Define
\begin{align*}
    \mathcal{E}_2\triangleq\{\exists(i,i)\in\mathcal{U}_1\times\mathcal{U}_2\text{ s.t. }W_{ii}\le y{|\mathcal{S}_1|}\}
\end{align*}
and 
\begin{align*}
    \mathcal{E}_3\triangleq\{\exists(i,j)\in\mathcal{U}_1\times\mathcal{U}_2\text{ s.t. }i\neq j\text{ and }W_{ij}>y{|\mathcal{S}_1|}\}.
\end{align*}
In the special case that $\mathcal{U}_1\times\mathcal{U}_2=\emptyset$, i.e., 
all the vertices are matched in Step 1, we set events $\mathcal{E}_2$ and $\mathcal{E}_3$ to be empty by convention and thus $\P(\mathcal{E}_2)=\P(\mathcal{E}_3)=0$. In the case that $\mathcal{U}_1\times\mathcal{U}_2\neq\emptyset$, event $\mathcal{E}_2^c\cap \mathcal{E}_3^c$ corresponds to the event that all non-anchor vertices are correctly matched through their user neighbors. 
\end{itemize}

We first show that  $\mathcal{E}_1^c\cap\mathcal{E}_2^c\cap\mathcal{E}_3^c$ implies that Algorithm~\AttrRich{} does not declare failure and it outputs $\hat{\pi}=\pi^*$.
Under event $\mathcal{E}_1^c$, it follows that there exists no conflicting pairs in $\mathcal{S}_1$, so Algorithm~\AttrRich{} does not declare failure at Step 1 and for each vertex $i\in\vuone\setminus\mathcal{U}_1$, we have $\hat{\pi}(i)=i$. 
Furthermore, $\mathcal{E}_1^c$ implies that $\mathcal{U}_1=\mathcal{U}_2$. Now we consider two different cases to complete the proof.
\begin{itemize}[leftmargin=1em]
    \item \underline{$\mathcal{U}_1=\mathcal{U}_2=\emptyset$:} In this case, all the vertices are correctly aligned through attribute neighbors. Therefore, Algorithm~\AttrRich{} terminates at Step 1 and outputs $\hat{\pi}=\pi^*$.
    \item \underline{$\mathcal{U}_1=\mathcal{U}_2\neq\emptyset$:} In this case, not all the vertices are aligned through attribute neighbors. Then event $\mathcal{E}_2^c\cap\mathcal{E}_3^c$ guarantees that for each $i\in\mathcal{U}_1$, we have 
    $$W_{ii}>(1-\delta_y)|\mathcal{S}_1|p\su^2$$
    and $$W_{ij}\le (1-\delta_y)|\mathcal{S}_1|p\su^2,\forall j\in\mathcal{U}_{2},i\neq j.$$
    Thus, the algorithm does not declare failure at Step 2 and we have for each $i\in\mathcal{U}_1$, $\hat{\pi}(i)=i$. Finally, it follows that $\hat{\pi}$ is a bijection and  $\hat{\pi}=\pi^*$.
\end{itemize}

Next, to prove Theorem~\ref{thm:achievability}, it suffices to show that $\P(\mathcal{E}_1^c\cap\mathcal{E}_2^c\cap\mathcal{E}_3^c)= 1-o(1)$.
However, both events $\mathcal{E}_2$ and $\mathcal{E}_3$ are based on two random sets $\mathcal{U}_1$ and $\mathcal{U}_2$. If we apply the union bound to upper bound the probability of $\mathcal{E}_2$ and $\mathcal{E}_3$ without any restriction on the size of $\mathcal{U}_1$ and $\mathcal{U}_2$, the bound will be very loose. 
Our key idea of the proof is to analyze the error events of Step~2 by conditioning on a carefully chosen event.  Specifically, let
the set of vertex pairs \emph{correctly matched} through attribute neighbors be denoted as
\[
\mathcal{S}_1'=\{(i,i)\in \mathcal{V}_1^\text{u}\times\mathcal{V}_2^\text{u}\text{ s.t. }C_{ii}>x\}.
\]
Note that $\mathcal{S}_1'\subseteq \mathcal{S}_1$ since $\mathcal{S}_1'$ only counts the correctly matched pairs. 
Then we consider the following auxiliary event:
\begin{align*}
    \mathcal{A}\triangleq\{n - |\mathcal{S}_1'| < n^c\},
\end{align*}
where $c\triangleq \max\left\{1-\frac{mqs_\text{a}^2}{(1+\epsilon/2)\log n},0\right\}$ (cf. the constant $\epsilon$ in Theorem~\ref{thm:achievability}).
To prove that $\P(\mathcal{E}_1^c\cap\mathcal{E}_2^c\cap\mathcal{E}_3^c)= 1-o(1)$, it then suffices to prove that $\P(\mathcal{A}\cap \mathcal{E}_1^c\cap\mathcal{E}_2^c\cap\mathcal{E}_3^c)= 1-o(1)$.
Notice that event $\mathcal{E}_1^c$ guarantees that the anchor set found in the first step only contains correctly matched pairs. It follows that event $\mathcal{E}_1^c$ implies $\mathcal{S}_1=\mathcal{S}_1'$. Therefore, event $\mathcal{A}\cap\mathcal{E}_1^c$ restricts the size of the set $\mathcal{U}_1\times\mathcal{U}_2$, so we can apply a much tighter union bound to upper bound the probability of $\mathcal{E}_2$ and $\mathcal{E}_3$ if we condition on the event $\mathcal{A}\cap\mathcal{E}_1^c$.

We now analyze  $\P(\mathcal{A}\cap \mathcal{E}_1^c\cap\mathcal{E}_2^c\cap\mathcal{E}_3^c)$.  Note that 
\begin{align}
    &\P(\mathcal{A}\cap\mathcal{E}_1^c\cap\mathcal{E}_2^c\cap\mathcal{E}_3^c)\nonumber\\
    &=\P(\mathcal{A}\cap\mathcal{E}_1^c)\P(\mathcal{E}_2^c\cap\mathcal{E}_3^c|\mathcal{A}\cap\mathcal{E}_1^c)\nonumber\\
    &=(1-\P(\mathcal{A}^c\cup\mathcal{E}_1))(1-\P(\mathcal{E}_2\cup\mathcal{E}_3|\mathcal{A}\cap\mathcal{E}_1^c))\nonumber\\
    &\ge(1-\P(\mathcal{A}^c)-\P(\mathcal{E}_1))(1-\P(\mathcal{E}_2|\mathcal{A}\cap\mathcal{E}_1^c)-\P(\mathcal{E}_3|\mathcal{A}\cap\mathcal{E}_1^c)).\label{thm1a}
\end{align}
where~\eqref{thm1a} follows by the union bound.
With Lemmas~\ref{lem:E1}--\ref{lem:E3} below, it is easy to see that $\P(\mathcal{A}\cap \mathcal{E}_1^c\cap\mathcal{E}_2^c\cap\mathcal{E}_3^c)= 1-o(1)$.

\begin{lemma}\label{lem:E1}
Consider $\mathcal{G}(n,p,s_\mathrm{u}; m, q,s_\mathrm{a})$ with $q=o(1)$ and $s_\mathrm{a}=\Theta(1)$. Assume that 
\[
mqs_\mathrm{a}^2=\Omega(\log n).
\]
Then
$\P(\mathcal{E}_1)=o(n^{-1/2}).$
\end{lemma}

\begin{lemma}\label{lem:A}

Consider $\mathcal{G}(n,p,s_\mathrm{u}; m, q,s_\mathrm{a})$ with $q=o(1)$ and $s_\mathrm{a}=\Theta(1)$. Assume that 
\[
mqs_\mathrm{a}^2=\Omega(\log n).
\]
Then
$\P(\mathcal{A})=1-o(1).$
\end{lemma}

\begin{lemma}\label{lem:E2}
Consider $\mathcal{G}(n,p,s_\mathrm{u}; m, q,s_\mathrm{a})$ with $p=o(1)$ and $s_\mathrm{u}=\Theta(1)$. Assume there exists some constant $\epsilon>0$ such that 
\begin{align}
    \label{eqLem3_00}
    mqs_\mathrm{a}^2+nps_\mathrm{u}^2 &\ge (1+\epsilon)\log n,\\
    \label{eqLem3_01}
    mqs_\mathrm{a}^2 &= \Omega(\log n).
\end{align}
Then $\P(\mathcal{E}_2|\mathcal{A}\cap \mathcal{E}_1^c)= {n^{-\Theta(1)}}$.
\end{lemma}
\begin{lemma}\label{lem:E3}
Consider $\mathcal{G}(n,p,s_\mathrm{u}; m, q,s_\mathrm{a})$ with $p=o(1)$ and $s_\mathrm{u}=\Theta(1)$. Assume there exists some constant $\epsilon>0$ such that 
\begin{align}
    \label{eqLem4_00}
    mqs_\mathrm{a}^2+nps_\mathrm{u}^2 &\ge (1+\epsilon)\log n,\\
    \label{eqLem4_01}
     mqs_\mathrm{a}^2 &= \Omega(\log n).
\end{align}
Then $\P(\mathcal{E}_3|\mathcal{A}\cap \mathcal{E}_1^c)={n^{-\Theta(1)}}$.
\end{lemma}

\section{Proof of Theorem~\ref{thm:achievability2}}\label{sec:proof-thm2}

Recall that Algorithm~\AttrSparse{} consists of the following two steps: In Step 1 (Align through attribute neighbors), the algorithm produces a set $\mathcal{S}_2$, which we refer to as the set of anchor pairs based on  $C_{ij}$ defined in~\eqref{eq:cij}; In Step 2 (Align through user neighbors), the algorithm performs two different processes depending on the sparsity of the user-user connections. In case when the user-user connection is sparse $np\le n^{1/7}$, the algorithm treats the anchors found in Step 1 as seeds and runs Algorithm~\ref{alg:Jiaming3}. In case the user-user connection is dense $np> n^{1/7}$, the algorithm explores the anchor vertices in the one-hop neighborhood of each non-anchor user vertex to align them. To prove Theorem~\ref{thm:achievability2}, we first consider two error events associated with Step 1, and then we separately consider the two different cases for Step 2.

Define 
\[
\mathcal{E}_4\triangleq\{\exists (i,j)\in\mathcal{V}_1^\text{u}\times\mathcal{V}_2^\text{u}\text{ s.t. }i\neq j\text{ and }C_{ij}\ge z\}.
\]
Event $\mathcal{E}_4^c$ guarantees the anchors found in Step 1 only contain correctly matched pairs. Let
\[
\mathcal{S}_2'=\{(i,i)\in \mathcal{V}_1^\text{u}\times\mathcal{V}_2^\text{u}\text{ s.t. }C_{ii}\ge z\}
\]
and define
\[
\mathcal{E}_5\triangleq\{|\mathcal{S}_2'|< n^{7/8}\}.
\]
Event $\mathcal{E}_4^c \cap\mathcal{E}_5^c$ guarantees the size of anchor set is large.
Given $\mathcal{E}_4^c \cap\mathcal{E}_5^c$, we first show that algorithm \AttrSparse{} outputs the correct permutation w.h.p., i.e, $\P(\hat{\Pi}\neq\Pi^*|\mathcal{E}_4^c\cap\mathcal{E}_5^c)=o(1)$. We consider two different cases for $np$.

\begin{itemize}[leftmargin=1em]
    
    \item \underline{$np\le n^{1/7}$.}  Recall that in this regime of $np$, Algorithm~\ref{alg:Jiaming3} is applied in Step 2. Notice that the choice of set $\mathcal{S}_2$ only depends on the user-attribute edges, so it is independent of all user-user edges. Therefore, by symmetry, when we run Algorithm~\ref{alg:Jiaming3}, we can view the seed set $\mathcal{I}_0$ as a set chosen uniformly at random over all subsets with size $|\mathcal{I}_0|$ and thus Theorem~\ref{cor:jiamingthm3} can directly be applied. So it suffices to show that the conditions in Theorem~\ref{cor:jiamingthm3} are satisfied under event $\mathcal{E}_4^c\cap\mathcal{E}_5^c$. Event $\mathcal{E}_4^c$ implies that $\mathcal{I}_0=\mathcal{S}_2=\mathcal{S}_2'$ and event $\mathcal{E}_5^c$ implies that $|\mathcal{I}_0|\ge n^{7/8}$.
    Moreover, because we assume that $np\su^2-\log n=\omega(1)$,
    we have $\P(\hat{\Pi}\neq\Pi^*|\mathcal{E}_4^c\cap\mathcal{E}_5^c)=o(1)$ by Theorem~\ref{cor:jiamingthm3} with $\beta=1/7$.
    
    
    \item \underline{$np>n^{1/7}$.} Recall that in this regime of $np$, we perform a similar process as Step 2 of Algorithm \AttrRich{}. We align the non-anchor vertices by exploring the anchors in their one-hop neighborhood. Similarly to the analysis for Theorem~\ref{thm:achievability}, we define two error events associated with this steps. Define
    \begin{align*}
    \mathcal{E}_6\triangleq\{\exists(i,i)\in\mathcal{U}_3\times\mathcal{U}_4\text{ s.t. }W_{ii}\le y|\mathcal{S}_2|\}
    \end{align*}
    and 
    \begin{align*}
    \mathcal{E}_7\triangleq\{\exists(i,j)\in\mathcal{U}_3\times\mathcal{U}_4\text{ s.t. }i\neq j\text{ and }W_{ij}>y|\mathcal{S}_2|\}.
    \end{align*}
    From the proof of Theorem~\ref{thm:achievability}, we know that $\{\hat{\Pi}\neq\Pi^*|\mathcal{E}_4^c\cap\mathcal{E}_5^c\}\subset\{\mathcal{E}_6^c\cap\mathcal{E}_7^c|\mathcal{E}_4^c\cap\mathcal{E}_5^c\}$. Therefore, we have
    \begin{align*}
        \P(\hat{\Pi}\neq\Pi^*|\mathcal{E}_4^c\cap\mathcal{E}_5^c)&\le \P(\mathcal{E}_6\cup\mathcal{E}_7|\mathcal{E}_4^c\cap\mathcal{E}_5^c)\\
        &\le \P(\mathcal{E}_6|\mathcal{E}_4^c\cap\mathcal{E}_5^c)+\P(\mathcal{E}_7|\mathcal{E}_4^c\cap\mathcal{E}_5^c).
    \end{align*}
    The statement $\P(\hat{\Pi}\neq\Pi^*|\mathcal{E}_4^c\cap\mathcal{E}_5^c)=o(1)$ follows from the Lemma below.
    \begin{lemma}\label{lem:E6}
Consider $\mathcal{G}(n,p,s_\mathrm{u}; m, q,s_\mathrm{a})$ with $p=o(1)$ and $s_\mathrm{u}=\Theta(1)$. Suppose that $np>n^{1/7}$.
Then we have $$\P(\mathcal{E}_6|\mathcal{E}_4^c\cap \mathcal{E}_5^c)=n^{-\Theta(1)}.$$
\end{lemma}
    \begin{lemma}\label{lem:E7}
Consider $\mathcal{G}(n,p,s_\mathrm{u}; m, q,s_\mathrm{a})$ with $p=o(1)$ and $s_\mathrm{u}=\Theta(1)$. Suppose that $np>n^{1/7}$.
Then we have $$\P(\mathcal{E}_7|\mathcal{E}_4^c\cap \mathcal{E}_5^c)= n^{-\Theta(1)}.$$
\end{lemma}
\end{itemize}
Finally, we have
\begin{align*}
    \P(\hat{\Pi}\neq\Pi^*)&=\P(\hat{\Pi}\neq\Pi^*,\mathcal{E}_4^c\cap\mathcal{E}_5^c)+\P(\hat{\Pi}\neq\Pi^*,\mathcal{E}_4\cup\mathcal{E}_5)\\
    &\le\P(\hat{\Pi}\neq\Pi^*|\mathcal{E}_4^c\cap\mathcal{E}_5^c)+\P(\mathcal{E}_4\cup\mathcal{E}_5)\\
    &\le o(1)+\P(\mathcal{E}_4)+\P(\mathcal{E}_5).
\end{align*}
With Lemmas~\ref{lem:E4} and~\ref{lem:E5} below, it is easy to see that $\P(\hat{\Pi}\neq\Pi^*)=o(1)$.
\begin{lemma}\label{lem:E4}
Consider $\mathcal{G}(n,p,s_\mathrm{u}; m, q,s_\mathrm{a})$ with $q=o(1)$ and $s_\mathrm{a}=\Theta(1)$. Assume that there exists some constant $\tau>0$ such that
\begin{equation}
    mq\sa^2\ge\frac{2\log n}{\tau\log \frac1q}.
    \label{eq:lemmaE4-cond}
    \end{equation}
    Then $\P(\mathcal{E}_4)=o(1)$.
\end{lemma}
\begin{lemma}\label{lem:E5}
Consider $\mathcal{G}(n,p,s_\mathrm{u}; m, q,s_\mathrm{a})$ with $q=o(1)$ and $s_\mathrm{a}=\Theta(1)$. Assume that
\begin{equation*}
    mq\sa^2=o(\log n).
    \end{equation*}
    Then $\P(\mathcal{E}_5)=o(1)$.
\end{lemma}
\section{Proof of Lemmas}

In this section, we prove the lemmas stated in the proofs of Theorems~\ref{thm:achievability} and~\ref{thm:achievability2}. To this end, we first state two technical lemmas that bound the binomial tail probability.

\begin{lemma}[Theorem 1 in~\cite{Hoeffding1963}]\label{lem:add_chernoff} 
Let $X\sim\mathrm{Binom}(n,\theta)$. Then we have:
\begin{itemize}
    \item $\P(X\ge n\theta+\lambda )\le \exp\left(-nD_\mathrm{KL}\left(\theta+\frac{\lambda}{n}||\theta\right)\right)$ for $0<\lambda<n-n\theta$;
    \item $\P(X\le n\theta-\lambda )\le \exp\left(-nD_\mathrm{KL}\left(\theta-\frac{\lambda}{n}||\theta\right)\right)$ for $0<\lambda<n\theta$,
   
\end{itemize}
 where $D_\mathrm{KL}(x||y)\triangleq x\log\frac{x}{y}+(1-x)\log\frac{1-x}{1-y}$ denotes the the Kullback--Leibler divergence between $\mathrm{Bern}(x)$ and $\mathrm{Bern}(y)$.
\end{lemma}
\begin{lemma}[Lemma 4.7.2 in~\cite{Ash1990}]\label{lem:tail_lower_bound}
Let $X\sim\mathrm{Binom}(n,\theta)$. Then we have
\[
\P(X\ge \lambda)\ge\frac{1}{\sqrt{8\lambda(1-\lambda/n)}}\exp\left\{-nD_\mathrm{KL}\left(\tfrac{\lambda}{n}\big|\big|\theta\right)\right\}
\]
for any $n\theta<\lambda\le n$.
\end{lemma}

The following lemma gives an approximation for the Kullback--Leibler divergence term in Lemmas~\ref{lem:add_chernoff} and~\ref{lem:tail_lower_bound}.

\begin{lemma}(Approximation of KL-divergence)
\label{lem:KLdiv}
Assume that {$\theta=o(1)$}, $s=\Theta(1)$, and $\Delta =\Theta(1)$.
Then we have
\begin{align}
    \label{eq6-1}
    D_\mathrm{KL}\left(\frac{\Delta}{\log{1/\theta}}\theta s^2 \bigg| \bigg| \theta s^2\right) &=  \theta s^2 + o(\theta),\\
    \label{eq6-2}
    D_\mathrm{KL}\left(\frac{\Delta}{\log{1/\theta}}\theta s^2 \bigg| \bigg| \theta^2s^2 \right) &= \Delta \theta s^2 + o(\theta).
\end{align}
\end{lemma}

With the three technical lemmas above, we are ready to prove Lemmas~\ref{lem:E1}-\ref{lem:E5}.

\subsection{Proof of Lemma~\ref{lem:E1}}
Recall that we defined error event $\mathcal{E}_1\triangleq\{\exists (i,j)\in\mathcal{V}_1^\text{u}\times\mathcal{V}_2^\text{u}\text{ s.t. }i\neq j\text{ and }C_{ij}>x\}.$ Here we prove $\P(\mathcal{E}_1)=o(n^{-1/2})$ under the assumption that $mq\sa^2=\Omega(\log n)$. To bound the probability of $\mathcal{E}_1$, we first consider distribution of random variable $C_{ij}$.
For two different vertices $i\in\vuone$ and $j\in\vutwo$ such that $i\neq j$, it follows from the definition of the model $\mathcal{G}(n,p,\su; m, q,\sa)$ that $C_{ij}\sim \text{Binom}(m,q^2\sa^2)$. Moreover, notice that 
$$x=\frac{\Delta_x}{\log\frac{1}{q}}mq\sa^2=\omega( mq^2\sa^2)=\omega(\E[C_{ij}]),$$
because $\Delta_x=\Theta(1)$ and $q=o(1)$. Now, we can upper bound the probability of error event $\mathcal{E}_1$ as
\begin{align}
    \P(\mathcal{E}_1)&=\P\{\exists (i,j)\in\mathcal{V}_1^\text{u}\times\mathcal{V}_2^\text{u}:i\neq j\text{ and }C_{ij}>x\}\nonumber\\
    &\le(n^2-n)\P\left(C_{12}>x\right)\label{lemma1-1}\\
    &\le n^2\exp\left(-mD_\text{KL}\left(\tfrac{x}{m}\big|\big|q^2\sa^2\right)\right)\label{lemma1-2}\\
    &=n^2\exp\left(-mD_\text{KL}\left(\frac{\Delta_x}{\log \frac{1}{q}}q\sa^2\bigg|\bigg|q^2\sa^2\right)\right)\nonumber\\
    &=n^2\exp(-m(\Delta_xq\sa^2+o(q)))\label{lemma1-3}\\
    &=\exp(2\log n-m(\Delta_xq\sa^2+o(q)))\nonumber\\
    &=o(n^{-1/2})\label{lemma1-4},
\end{align}
where~\eqref{lemma1-1} follows from the union bound,~\eqref{lemma1-2} follows from Lemma~\ref{lem:add_chernoff},~\eqref{lemma1-3} follows from Lemma~\ref{lem:KLdiv} and~\eqref{lemma1-4} follows since constant $\Delta_x\ge \frac{3\log n}{mq\sa^2}$.
\subsection{Proof of Lemma~\ref{lem:A}}

Recall that we defined auxiliary event $\mathcal{A}\triangleq\{n - |\mathcal{S}_1'| < n^c\}$, where $\mathcal{S}_1'=\{(i,i)\in \mathcal{V}_1^\text{u}\times\mathcal{V}_2^\text{u}\text{ s.t. }C_{ii}>x\}$. Here we prove $\P(\mathcal{A})=1-o(1)$ under the assumption that $mq\sa^2=\Omega(\log n)$. To show this, we first consider the distribution of random variable $C_{ii}$ and upper bound the probability of event $\{C_{ii}\le x\}$.
For each vertex $i\in\vuone$, it follows from the definition of the model $\mathcal{G}(n,p,\su; m, q,\sa)$ that $C_{ii}\sim\text{Binom}(m,q\sa^2)$. Notice that 
\[
x=\frac{\Delta_x}{\log\frac{1}{q}}mq\sa^2=o( mq\sa^2)=o(\E[C_{ii}]),
\]
because $q=o(1)$ and $\Delta_x=\Theta(1)$. We can upper bound the probability of the tail event $\{C_{ii}\le x\}$ using Lemma~\ref{lem:add_chernoff}:
\begin{align}
    \P(C_{ii}\le x)&\le \exp\left(-mD_\text{KL}\left(\frac{x}{m}||q\sa^2\right)\right)\nonumber\\ 
    &=\exp\left(-mD_\text{KL}\left(\frac{\Delta_x}{\log\frac{1}{q}}q\sa^2||q\sa^2\right)\right)\nonumber\\ 
    &=\exp\left(-m\left(q\sa^2+o(q)\right)\right)\label{eq:not_in_S_upper_bound},
\end{align}
where~\eqref{eq:not_in_S_upper_bound} follows from Lemma~\ref{lem:KLdiv}. For simplicity, we denote $\exp\left(-mD_\text{KL}\left(\frac{\Delta_x}{\log\frac{1}{q}}q\sa^2||q\sa^2\right)\right)$ by $\gamma$ from this point. By~\eqref{eq:not_in_S_upper_bound} and the assumption that $mq\sa^2=\Omega(\log n)$, we have $\gamma=o(1)$.

Furthermore, notice that for each different $i\in\vuone$, the random variable $C_{ii}$ are independent and identically distributed. Therefore, the number of vertices $i\in\vuone$ with $C_{ii} \le x$ is distributed according to the Binomial distribution
$$n-|\mathcal{S}_1'|\sim \text{Binom}(n,\P(C_{ii}\le x)).$$
By the upper bound~\eqref{eq:not_in_S_upper_bound}, for any positive number $z$, we have
\begin{align}
\P(n-|\mathcal{S}_1'|\ge z)\le\P\left(\text{Binom}\left(n,\gamma\right)\ge z\right).\label{eq:domination}
\end{align}
Recall that $c\triangleq \max\left\{1-\frac{mq\sa^2}{(1+\epsilon/2)\log n},0\right\}$, where $\epsilon$ is a positive constant. We have
\begin{align}
    \frac{n^{c-1}}{\gamma}&\ge n^{-\frac{mq\sa^2}{(1+\epsilon/2)\log n}}\exp\left(mD_\text{KL}\left(\frac{\Delta_x}{\log\frac{1}{q}}q\sa^2||q\sa^2\right)\right)\nonumber\\
    &=\exp\left(-\frac{mq\sa^2}{1+\epsilon/2}+mq\sa^2+o(mq)\right)\label{lemma2b}\\
    &=\omega(1)\label{lemma2c1},
\end{align}
where~\eqref{lemma2b} follows from~\eqref{eq:not_in_S_upper_bound}, and~\eqref{lemma2c1} follows since $mq\sa^2=\Omega(\log n)$.
Finally, we can upper bound the probability of event $\mathcal{A}^c$ as
\begin{align}
    &\P(\mathcal{A}^c)\nonumber\\&=\P(n-|\mathcal{S}_1'|\ge n^c)\nonumber\\
    &\le\P\left(\text{Binom}\left(n,\gamma\right)\ge n^{c}\right)\label{lemma2c}\\
    &\le\exp(-nD_\text{KL}(n^{c-1}||\gamma))\label{lemma2d}\\
    &=\exp\left(-n\left(n^{c-1}\log\tfrac{n^{c-1}}{\gamma}+(1-n^{c-1})\log\tfrac{1-n^{c-1}}{1-\gamma}\right)\right)\nonumber\\
    &=\exp\bigg(-n\bigg(n^{c-1}\log\tfrac{n^{c-1}}{\gamma}\nonumber\\
    &\;\;\;\;+(1-n^{c-1})\tfrac{\gamma-n^{c-1}}{1-\gamma}(1+o(1))\bigg)\bigg)\label{lemma2e}\\
    &=\exp\left(-n\left(\omega(n^{c-1})-n^{c-1}(1+o(1))\right)\right)\label{lemma2f}\\
    &=\exp\left(-n\cdot \omega(n^{c-1})\right)\nonumber\\
    &=o(1)\label{lemma2g},
\end{align}
where~\eqref{lemma2c} follows from~\eqref{eq:domination},~\eqref{lemma2d} follows from Lemma~\ref{lem:add_chernoff},~\eqref{lemma2e} follow from the Taylor expansion of the function $\log x$ and the fact that $n^{c-1}=o(1)$ and $\gamma=o(1)$,~\eqref{lemma2f} follows since $n^{c-1}=\omega(\gamma)$ and finally,~\eqref{lemma2g} follows from the fact that $c\ge 0$.
\subsection{Proof of Lemma~\ref{lem:E2}}
Here we prove that the conditions \eqref{eqLem3_00} and \eqref{eqLem3_01} imply $\prob(\mathcal{E}_2 |\mathcal{A} \cap \mathcal{E}_1^c ) =n^{-\Theta(1)}$. Recall that we defined events
\begin{align*}
    \mathcal{E}_2 &= \{\exists(i,i)\in\mathcal{U}_1\times\mathcal{U}_2:W_{ii}\le y{|\mathcal{S}_1|}\},\\
    \mathcal{A} &= \{n - |\mathcal{S}_1'| < n^c\},\\
    \mathcal{E}_1 &= \{\exists (i,j)\in\mathcal{V}_1^\text{u}\times\mathcal{V}_2^\text{u}\text{ s.t. }i\neq j\text{ and }C_{ij}>x\}.
\end{align*}
To analyze the conditional event $\mathcal{E}_2 |\mathcal{A} \cap \mathcal{E}_1^c$, first notice that  $\mathcal{E}_1^c$ implies that only identical pairs are in the anchor set, i.e., $\mathcal{U}_1 =\mathcal{U}_2 $. 
Thus, we obtain a simplified expression of the conditional event as
\[
\mathcal{E}_2 |\mathcal{A} \cap \mathcal{E}_1^c = 
\{ \exists i\in \mathcal{U}_1: W_{ii}\leq y{|\mathcal{S}_1|}\} \mid \mathcal{A} \cap \mathcal{E}_1^c. 
\]
The condition on the auxiliary event  $\mathcal{A}$ further implies that the number of identical pairs that are not discovered in the anchor set is at most $n^c$, i.e., $|\mathcal{U}_1|<n^c$.
Here, recall that $c = \max \left\{ 1-\frac{mqs_\text{a}^2}{(1+\epsilon/2)\log n}, 0 \right\}$. We note that $c=0$ implies that the unmatched users set $\mathcal{U}_1= \emptyset$, which is taken care of by a separate analysis in the proof of Theorem~\ref{thm:achievability}. For the remaining analysis in this Lemma, we only consider the case where $\mathcal{U}_1\neq \emptyset$, and consequently we have $c= 1-\frac{mqs_\text{a}^2}{(1+\epsilon/2)\log n} $. 

Applying the union bound on the conditional error event, we have 
\begin{align}
    &\prob(\mathcal{E}_2 |\mathcal{A} \cap \mathcal{E}_1^c) = \prob(\exists i\in \mathcal{U}_1,W_{ii}\leq y{|\mathcal{S}_1|} \mid \mathcal{A} \cap \mathcal{E}_1^c )\nonumber\\
    \label{eq3-0}
    &=\sum_{k=0}^{n^c} 
    \prob(\exists i\in \mathcal{U}_1,W_{ii}\leq y{|\mathcal{S}_1|} \mid |\mathcal{U}_1|=k, \mathcal{E}_1^c ) 
    \prob(|\mathcal{U}_1|=k \mid \mathcal{A}\cap \mathcal{E}_1^c)  \\
    \label{eq3-1}
    & \leq \sum_{k=0}^{n^c}
    k \prob(W_{11}\leq y{|\mathcal{S}_1|} \mid |\mathcal{U}_1|=k, \mathcal{E}_1^c)
    \prob(|\mathcal{U}_1|=k \mid\mathcal{A}\cap \mathcal{E}_1^c)\\
    \label{eq3-2}
    & \leq \max_{k \in [0,n^c]}\{k  \prob(W_{11}\leq y{|\mathcal{S}_1|} \mid |\mathcal{U}_1|=k, \mathcal{E}_1^c) \}.
\end{align}
{In \eqref{eq3-0}, we have $n^c$ as the upper limit in the summation because of conditioning on $\mathcal{A}$. 
Equation \eqref{eq3-1} follows from the union bound.}

Next, we upper bound $\prob(W_{11}\leq y{|\mathcal{S}_1|} \mid |\mathcal{U}_1|=k, \mathcal{E}_1^c)$.  
Recall

$$W_{11}=\sum_{v\in \mathcal{N}_1^\text{u}(1), u\in \mathcal{N}_2^\text{u}(1)}\indi_{\{(v,u)\in \mathcal{S}_1\}}$$
counts the number of aligned anchor neighbors of a user vertex.
To see the conditional distribution of $W_{11}$, notice that conditioned on events $\mathcal{E}_1^c$ and $|\mathcal{U}_1| = k$, the whole anchor set $\mathcal{S}_1$ only contains identical pairs and is of size $n-k$.
Thus, we get the following simpler expression 
\begin{align*}
    W_{11} \mid \{|\mathcal{U}_1|=k, \mathcal{E}_1^c\} &=\sum_{v\in \mathcal{N}_1^\text{u}(1), v\in \mathcal{N}_2^\text{u}(1)}\indi_{\{(v,v)\in \mathcal{S}_1\}}\\
    &= \sum_{(v,v)\in \mathcal{S}_1} \indi\{v \in \mathcal{N}_1^\text{u}(1), v\in \mathcal{N}_2^\text{u}(1) \}.
\end{align*}
{Here this random variable $W_{11} \mid \{|\mathcal{U}_1|=k, \mathcal{E}_1^c\}$ is the summation of $|\mathcal{S}_1|=n-k$ independent and identically distributed Bernoulli random variables. Each Bernoulli random variable takes value 1 when a pair of anchor vertices 
connect to vertex 1 in both $G_1$ and $G_2$, and this happens with probability $p\su^2$.
We therefore have $W_{11} \mid \{|\mathcal{U}_1|=k, \mathcal{E}_1^c\} \sim \Bin(n-k, p\su^2)$. To upper bound the probability $\prob(W_{11}\leq y{|\mathcal{S}_1|} \mid |\mathcal{U}_1|=k, \mathcal{E}_1^c)$ in \eqref{eq3-2}, we use the Chernoff bound from Lemma~\ref{lem:add_chernoff} and get }
\begin{align}
    &\prob(W_{11}\leq y{|\mathcal{S}_1|} \mid |\mathcal{U}_1|=k, \mathcal{E}_1^c) \nonumber \\
    \label{eq3-3}
    &\leq \exp\left\{-(n-k) \DKL\left(\frac{\Delta_y}{\log 1/p}p\su^2 || p\su^2 \right) \right\}.
\end{align}

Plugging \eqref{eq3-3} into the previous \eqref{eq3-2}, we finally have 
\begin{align}
    &\prob(\mathcal{E}_2 |\mathcal{A} \cap \mathcal{E}_1^c) \nonumber \\
    &\leq \max_{k \in [0,n^c]}\left\{k \exp\{-(n-k) \DKL\left(\frac{\Delta_y}{\log 1/p}p\su^2 || p\su^2 \right)  \right\} \nonumber \\
    &= n^c \exp \left\{-(n-n^c) \DKL\left(\frac{\Delta_y}{\log 1/p}p\su^2 || p\su^2 \right) \right\}\nonumber \\
    \label{eq3-4}
    &= \exp \{ c\log n-(n-n^c) (p\su^2 + o(p\su^2))\}\\
    \label{eq3-5}
    & \leq \exp\left\{ \log n - \frac{mqs_\text{a}^2}{(1+\epsilon/2)}  -np\su^2 +o(np)\right\}\\
    \label{eq3-6}
    & \leq \exp\left\{ \log n - \frac{mqs_\text{a}^2 + np\su^2}{(1+\epsilon/2)} \right\}\\
    \label{eq3-7}
    & \leq \exp \left\{-\frac{\epsilon/2 }{1+\epsilon/2 }\log n \right\}\\
    & ={n^{-\Theta(1)}}\nonumber.
\end{align}
Here \eqref{eq3-4} follows from the KL-divergence approximation \eqref{eq6-1} in Lemma~\ref{lem:KLdiv}.
We get \eqref{eq3-5} by plugging in $c=1-\frac{mqs_\text{a}^2}{(1+\epsilon/2)\log n} $ and applying assumption \eqref{eqLem3_01} $mqs_\mathrm{a}^2 = \Omega(\log n)$. 
Equation \eqref{eq3-6} follows because {$o(np) \leq \frac{\epsilon/2}{1+\epsilon/2} np\su^2 =\Theta(np)$.}
Equation \eqref{eq3-7} follows from the condition~\eqref{eqLem3_00}, which requires that  $mqs_\mathrm{a}^2+nps_\mathrm{u}^2 \ge (1+\epsilon)\log n$ for a constant $\epsilon$.
\subsection{Proof of Lemma~\ref{lem:E3}}

Here we prove that conditions \eqref{eqLem4_00} and \eqref{eqLem4_01} imply $\prob(\mathcal{E}_3 |\mathcal{A} \cap \mathcal{E}_1^c ) =n^{-\Theta(1)}$. The conditioned events here are exactly the same as those of Lemma~\ref{lem:E2}, and we use a similar proof strategy. Recall that we defined the event 
\begin{align*}
    \mathcal{E}_3\triangleq\{\exists(i,j)\in\mathcal{U}_1\times\mathcal{U}_2\text{ s.t. }i\neq j\text{ and }W_{ij}>y{|\mathcal{S}_1|}\}.
\end{align*}
To analyze  the  conditional event $\mathcal{E}_3 |\mathcal{A} \cap \mathcal{E}_1^c$, we reuse the same observation from the proof of Lemma~\ref{lem:E2}, that only identical pairs are in the anchor set. Thus, we are able to simplify the expression as
\[
\mathcal{E}_3\mid \mathcal{A} \cap \mathcal{E}_1^c = \{\exists i, j \in \mathcal{U}_1, i\neq j: W_{ij}>y{|\mathcal{S}_1|}\}\mid \mathcal{A} \cap \mathcal{E}_1^c,
\]
where the number of unaligned identical pairs $|\mathcal{U}_1| <n^c$. Here, we also have $c= 1-\frac{mqs_\text{a}^2}{(1+\epsilon/2)\log n} $, because $c=0$ implies that the unmatched users set $\mathcal{U}_1= \emptyset$, which is taken care of by a separate analysis in the proof of Theorem~\ref{thm:achievability}.

Applying the union bound on the conditional error event, we get
\begin{align}
    &\prob(\mathcal{E}_3 |\mathcal{A} \cap \mathcal{E}_1^c) = \prob(\exists i, j \in \mathcal{U}_1, i\neq j, W_{ij} >y{|\mathcal{S}_1|} \mid \mathcal{A} \cap \mathcal{E}_1^c )\nonumber\\
    &=\sum_{k=0}^{n^c} 
    \prob(\exists i, j \in \mathcal{U}_1, i\neq j, W_{ij} > y{|\mathcal{S}_1|} \mid |\mathcal{U}_1|=k, \mathcal{E}_1^c ) \nonumber\\
    & \hspace{36pt} \times \prob(|\mathcal{U}_1|=k \mid \mathcal{A} \cap \mathcal{E}_1^c) \nonumber \\
    \label{eq4-1}
    &\leq \sum_{k=0}^{n^c} k^2 \prob( W_{12}> y{|\mathcal{S}_1|} \mid |\mathcal{U}_1|=k, \mathcal{E}_1^c )
    \prob(|\mathcal{U}_1|=k \mid \mathcal{A} \cap \mathcal{E}_1^c)\\
    \label{eq4-2}
    & \leq \max_{k \in [0,n^c]} \{k^2 \prob( W_{12}> y{|\mathcal{S}_1|} \mid |\mathcal{U}_1|=k, \mathcal{E}_1^c )\large\}.
\end{align}

{Here we further upper bound $\prob(W_{12} > y{|\mathcal{S}_1|} \mid {|\mathcal{U}_1|=k}, \mathcal{E}_1^c)$ in \eqref{eq4-2}.
Recall that 
\[W_{12}=\sum_{v\in \mathcal{N}_1^\text{u}(1), u\in \mathcal{N}_2^\text{u}(2)}\indi_{\{(v,u)\in \mathcal{S}_1\}},\]
represents the number of aligned anchor neighbors of user vertex $1$ in $G_1$ and user vertex $2$ in $G_2$,
Notice that conditioned on events  $\mathcal{E}_1^c$ and $|\mathcal{U}_1| = k$ ,  the anchor set $\mathcal{S}_1$ only contains identical pairs and is of size $n-k$. 
Thus, we get the following simpler expression }
\begin{align*}
    W_{12} \mid \{{|\mathcal{U}_1|=k}, \mathcal{E}_1^c\} &=\sum_{v\in \mathcal{N}_1^\text{u}(1), v\in \mathcal{N}_2^\text{u}(2)}\indi_{\{(v,v)\in \mathcal{S}_1\}}\\
    &= \sum_{(v,v)\in \mathcal{S}_1} \indi\{v \in \mathcal{N}_1^\text{u}(1), v\in \mathcal{N}_2^\text{u}(2) \}
\end{align*}
{Here the random variable $W_{12} \mid \{{|\mathcal{U}_1|=k}, \mathcal{E}_1^c\}$ is the summation of $n-k$ independent and identically distributed Bernoulli random variables. Each Bernoulli random variable takes value 1 when a pair of anchor vertices in  $\mathcal{S}_1$ connect to vertex 1 in $G_1$ and vertex 2 in $G_2$ and this happens with probability $p^2\su^2$.
Therefore, we have $W_{12} | \{|\mathcal{U}_1|=k, \mathcal{E}_1^c\} \sim \Bin(n-k, p^2\su^2)$. We then apply Chernoff bound in Lemma~\ref{lem:add_chernoff} and get }
\begin{align}
    &\prob(W_{12}>y{|\mathcal{S}_1|} \mid |\mathcal{U}_1|=k, \mathcal{E}_1^c) \nonumber \\
    \label{eq4-3}
    &\leq \exp\left\{-(n-k) \DKL\left(\frac{\Delta_y}{\log 1/p}p\su^2 || p^2\su^2 \right)  \right\}
\end{align}
Plugging \eqref{eq4-3} into \eqref{eq4-2} , we have 
\begin{align}
    \nonumber
     &\prob(\mathcal{E}_3 |\mathcal{A} \cap \mathcal{E}_1^c) \\
     \nonumber
     & \leq \max_{k \in [0,n^c]} \left\{k^2 \exp\{-(n-k) \DKL\left(\frac{\Delta_y}{\log 1/p}p\su^2 || p^2\su^2 \right)  \right\}\\
     \nonumber
     &= n^{2c} \exp \left\{-(n-n^c) \DKL\left(\frac{\Delta_y}{\log 1/p}p\su^2 || p^2\su^2 \right) \right\} \\
     \label{eq4-4}
     & \leq \exp\{2c\log n -(n-n^c) ( \Delta_y p\su^2 + o(p))\}\\
     \label{eq4-5}
     & \leq \exp \left\{ 2\log n - \frac{2mq\sa^2}{1+\epsilon/2} - \Delta_y np\su^2 + o(pn) \right\}\\
     \label{eq4-6}
     & \leq \exp \left\{ 2\log n - \frac{2mq\sa^2 +2np\su^2}{1+\epsilon/2} \right\}\\
     \label{eq4-7}
     & \leq \exp \left\{-\frac{\epsilon }{1+\epsilon/2 }\log n \right\}\\
     & = n^{-\Theta(1)}.
\end{align}
Here \eqref{eq4-4} follows from the KL-divergence approximation formula \eqref{eq6-2} in Lemma~\ref{lem:KLdiv}.
We get \eqref{eq4-5} by plugging in $c=1-\frac{mqs_\text{a}^2}{(1+\epsilon/2)\log n} $ and applying assumption \eqref{eqLem4_01} $mqs_\mathrm{a}^2 = \Omega(\log n)$. Equation \eqref{eq4-6} follows from the fact that { $o(np) \leq \frac{\epsilon/2 \Delta_y }{1+\epsilon/2} np\su^2=\Theta(np)$}, and condition that $\Delta_y \geq 2$. Equation \eqref{eq4-7} follows from the condition~\eqref{eqLem4_00}.
\subsection{Proof of Lemma~\ref{lem:E6}}
Recall that we defined events
\begin{align*}
    &\mathcal{E}_4=\{\exists (i,j)\in\mathcal{V}_1^\text{u}\times\mathcal{V}_2^\text{u}\text{ s.t. }i\neq j\text{ and }C_{ij}\ge z\},\\
    &\mathcal{E}_5=\{|\mathcal{S}_2'|< n^{7/8}\},\\
    &\mathcal{E}_6=\{\exists(i,i)\in\mathcal{U}_3\times\mathcal{U}_4\text{ s.t. }W_{ii}\le y|\mathcal{S}_2|\}.
\end{align*}
Here we prove that $\P(\mathcal{E}_6|\mathcal{E}_4^c\cap \mathcal{E}_5^c)= o(1)$ under the assumption that $np>n^{1/7}$. 
By the analogous argument as for equations~\eqref{eq3-2} and~\eqref{eq3-3} respectively, we have that 
\begin{align}
   &\P(\mathcal{E}_6|\mathcal{E}_4^c\cap \mathcal{E}_5^c)\nonumber\\
   &\le  \max_{k \in [0,n-n^{7/8}]}\{k  \prob(W_{11}\leq y|\mathcal{S}_2| \mid |\mathcal{U}_3|=k, \mathcal{E}_4^c) \}\label{eq:E6_1},
\end{align}
and
\begin{align}
    &\prob(W_{11}\leq y|\mathcal{S}_2| \mid |\mathcal{U}_3|=k, \mathcal{E}_4^c)\nonumber\\
    &\le \exp\left\{-(n-k) \DKL\left(\frac{\Delta_y}{\log 1/p}p\su^2 || p\su^2 \right) \right\}\label{eq:E6_2}.
\end{align}
Finally, plugging equation~\eqref{eq:E6_2} into equation~\eqref{eq:E6_1} gives
\begin{align}
    &\P(\mathcal{E}_6|\mathcal{E}_4^c\cap \mathcal{E}_5^c)\nonumber\\
    &\le  \max_{k \in [0,n-n^{7/8}]}\left\{k  \exp\left\{-(n-k) \DKL\left(\tfrac{\Delta_y}{\log 1/p}p\su^2 || p\su^2 \right) \right\} \right\}\nonumber\\
    &\le n\exp\left(-n^{7/8}\DKL\left(\tfrac{\Delta_y}{\log 1/p}p\su^2 || p\su^2 \right)\right)\nonumber\\
    &\le \exp(\log n-n^{7/8} (p\su^2 + o(p\su^2)))\label{eq:E6_3}\\
    &=n^{-\Theta(1)}\label{eq:E6_4},
\end{align}
where~\eqref{eq:E6_3} follows from the KL-divergence approximation \eqref{eq6-1} in Lemma~\ref{lem:KLdiv} and~\eqref{eq:E6_4} follows because $np>n^{1/7}$.
\subsection{Proof of Lemma~\ref{lem:E7}}
Recall that we defined event 
\begin{align*}
    \mathcal{E}_7\triangleq\{\exists(i,j)\in\mathcal{U}_3\times\mathcal{U}_4\text{ s.t. }i\neq j\text{ and }W_{ij}>y|\mathcal{S}_2|\}.
\end{align*}
Here we prove that $\P(\mathcal{E}_7|\mathcal{E}_4^c\cap \mathcal{E}_5^c)= o(1)$ under the assumption that $np>n^{1/7}$. By the analogous argument as for equations~\eqref{eq4-2} and~\eqref{eq4-3}, we have that 
\begin{align}
   &\P(\mathcal{E}_7|\mathcal{E}_4^c\cap \mathcal{E}_5^c)\nonumber\\
   &\le  \max_{k \in [0,n-n^{7/8}]}\{k^2  \prob(W_{12}> y|\mathcal{S}_2| \mid |\mathcal{U}_3|=k, \mathcal{E}_4^c) \}\label{eq:E7_1},
\end{align}
and
\begin{align}
    &\prob(W_{12}> y|\mathcal{S}_2| \mid |\mathcal{U}_3|=k, \mathcal{E}_4^c)\nonumber\\
    &\le \exp\left\{-(n-k) \DKL\left(\frac{\Delta_y}{\log 1/p}p\su^2 || p^2\su^2 \right) \right\}.
    \label{eq:E7_2}
\end{align}
Plugging equation~\eqref{eq:E7_2} into equation~\eqref{eq:E7_1} gives
\begin{align}
    &\P(\mathcal{E}_7|\mathcal{E}_4^c\cap \mathcal{E}_5^c)\nonumber\\
    &\le  \max_{k \in [0,n-n^{7/8}]}\left\{k^2  \exp\left\{-(n-k) \DKL\left(\tfrac{\Delta_y}{\log 1/p}p\su^2 || p^2\su^2 \right) \right\} \right\}\nonumber\\
    &\le n^2\exp\left(-n^{7/8}\DKL\left(\tfrac{\Delta_y}{\log 1/p}p\su^2 || p^2\su^2 \right)\right)\nonumber\\
    &\le \exp(2\log n-n^{7/8} (\Delta_y p\su^2 + o(p))\label{eq:E7_3}\\
    &=o(1)\label{eq:E7_4},
\end{align}
where \eqref{eq:E7_3} follows from the KL-divergence approximation formula \eqref{eq6-2} in Lemma~\ref{lem:KLdiv} and~\eqref{eq:E7_4} follows because $np>n^{1/7}$ and $\Delta_y=\Theta(1)$.
\subsection{Proof of Lemma~\ref{lem:E4}}
Recall that we defined error event $\mathcal{E}_4\triangleq\{\exists (i,j)\in\mathcal{V}_1^\text{u}\times\mathcal{V}_2^\text{u}\text{ s.t. }i\neq j\text{ and }C_{ij}\ge z\}$. Here we prove $\P(\mathcal{E}_4)=o(1)$ under the assumption that $mq\sa^2\ge\frac{2\log n}{\tau\log\frac1q}$. To bound the probability of $\mathcal{E}_4$, we first consider the distribution of random variable $C_{ij}$.
For two different vertices $i\in\vuone$ and $j\in\vutwo$ such that $i\neq j$, it follows from the definition of the model $\mathcal{G}(n,p,s_u; m, q,s_a)$ that $C_{ij}\sim \text{Binom}(m,q^2s_\text{a}^2)$. Moreover, notice that 
$$z= (1+\tau)mqs_\text{a}^2=\omega( mq^2s_\text{a}^2)=\omega(\E[C_{ij}]),$$
since $q=o(1)$. Therefore, we can upper bound the probability of error event $\mathcal{E}_4$ as
\begin{align}
    &\P(\mathcal{E}_4)\nonumber\\
    &=\P\{\exists (i,j)\in\mathcal{V}_1^\text{u}\times\mathcal{V}_2^\text{u}:i\neq j\text{ and }C_{ij}\ge z\}\nonumber\\
    &\le(n^2-n)\P\left(C_{12}\ge z\right)\label{eq:lemmaE4-1}\\
    &\le n^2\exp\left(-mD_\text{KL}\left(\tfrac{z}{m}\big|\big|q^2s_\text{a}^2\right)\right)\label{eq:lemmaE4-2}\\
    &=n^2\exp\left(-mD_\text{KL}\left((1+\tau)qs_\text{a}^2||q^2s_\text{a}^2\right)\right)\nonumber\\
    &=n^2\exp\left(-m(1+\tau)q\sa^2\log\frac{1+\tau}{q}\right)\nonumber\\
    &\;\;\;\cdot\exp\left(-m(1-(1+\tau)q\sa^2)\log\frac{1-(1+\tau)q\sa^2}{1-q^2\sa^2}\right)\nonumber\\
    &=n^2\exp\left(-m(1+\tau)q\sa^2\log\frac{1+\tau}{q}\right)\nonumber\\
     &\;\;\;\cdot\exp\left((1+o(1))m(1-(1+\tau)q\sa^2)\frac{(1+\tau) q\sa^2-q^2\sa^2}{1-q^2\sa^2}\right)\nonumber\\
    &=n^2\exp\left(-m(1+\tau)q\sa^2\log\frac{1+\tau}{q}\right)\nonumber\\
     &\;\;\;\cdot\exp\left((1+o(1))m(1+\tau)q\sa^2\right)\nonumber\\
    &=\exp\left(2\log n-m(1-o(1))(1+\tau)q\sa^2\log\frac{1}{q}\right)\nonumber\\
    &=\exp(-\Omega(\log n))\label{eq:lemmaE4-3}\\
    &=n^{-\Omega(1)},\nonumber
\end{align}
where~\eqref{eq:lemmaE4-1} follows by the union bound,~\eqref{eq:lemmaE4-2} follows from Lemma~\ref{lem:add_chernoff}, and~\eqref{eq:lemmaE4-3} follows since in~\eqref{eq:ac2_cond4} we assume that $mq\log\frac{1}{q}\tau\sa^2\ge 2\log n$ and $\tau=\Theta(1)$.

\subsection{Proof of Lemma~\ref{lem:E5}}
Recall that we defined error event $\mathcal{E}_5\triangleq\{|\mathcal{S}_2'|< n^{7/8}\}$, where $\mathcal{S}_2'=\{(i,i)\in \mathcal{V}_1^\text{u}\times\mathcal{V}_2^\text{u}\text{ s.t. }C_{ii}\ge z\}$. Here we prove $\P(\mathcal{E}_5)=o(1)$ under the assumption that $mq\sa^2=o(\log n)$. To show this, we first consider the distribution of random variable $C_{ii}$ and lower bound the probability of event $\{C_{ii}\ge z\}$.
For each vertex $i\in\vuone$, it follows from the definition of the model $\mathcal{G}(n,p,\su; m, q,\sa)$ that $C_{ii}\sim\text{Binom}(m,qs_\text{a}^2)$. Notice that 
\[
z=(1+\tau)mqs_\text{a}^2> \E[C_{ii}].
\]
We can lower bound the probability of the tail event $\{C_{ii}\ge z\}$ using Lemma~\ref{lem:tail_lower_bound} as
\begin{align}
    &\P(C_{ii}\ge z)\nonumber\\
    &\ge \frac{1}{\sqrt{8z(1-z/m)}}\exp\left(-mD_\mathrm{KL}\left(\tfrac{z}{m}\big|\big|q\sa^2\right)\right)\nonumber\\ 
    &\ge \frac{1}{\sqrt{8z}}\exp\left(-mD_\mathrm{KL}\left((1+\tau)q\sa^2||q\sa^2\right)\right)\nonumber\\ 
    &=\exp\left(-\frac{1}{2}\log(8(1+\tau)mq\sa^2)\right)\nonumber\\
    &\;\;\;\cdot\exp(-m(1+\tau)q\sa^2\log(1+\tau))\nonumber\\
    &\;\;\;\cdot\exp\left(-m(1-(1+\tau)q\sa^2)\log\frac{1-(1+\tau)q\sa^2}{1-q\sa^2}\right)\nonumber\\
    &=\exp\left(-\frac{1}{2}\log(8(1+\tau)mq\sa^2)\right)\nonumber\\
    &\;\;\;\cdot\exp(-m(1+\tau)q\sa^2\log(1+\tau))\nonumber\\
    &\;\;\;\cdot\exp\left((1+o(1))(1-(1+\tau)q\sa^2)m\frac{\tau q\sa^2}{1-q\sa^2}\right)\nonumber\\
    &=\exp\left(-\frac{1}{2}\log(8(1+\tau)mq\sa^2)\right)\nonumber\\
    &\;\;\;\cdot\exp(-(1+o(1))((1+\tau)\log(1+\tau)-\tau)mq\sa^2)\nonumber\\
    &=\exp(-o(\log n))\label{eq:lemmaE5-1}\\
    &\ge n^{-1/9}\label{eq:lemmaE5-2},
\end{align}
where~\eqref{eq:lemmaE5-1} follows since $mq\sa^2=o(\log n)$ and that $\tau=\Theta(1)$.
Furthermore, notice that for each different $i\in\vuone$, the random variable $C_{ii}$ are independent and identically distributed. Therefore, the number of vertices $i\in\vuone$ with $C_{ii} \ge z$ is distributed according to the Binomial distribution
$$|\mathcal{S}_2'|\sim \text{Binom}(n,\P(C_{ii}\ge z)).$$
By the lower bound~\eqref{eq:lemmaE5-2}, for any positive number $w$, we have
\begin{align}
\P(|\mathcal{S}_2'|\le w)\le\P\left(\text{Binom}\left(n,n^{-1/9}\right)\le w\right).\label{eq:lemmaE5-3}
\end{align}
Finally we can upper bound the probability of event $\mathcal{E}_5$ as
\begin{align}
    &\P(\mathcal{E}_5)\nonumber\\
    &=\P(|\mathcal{S}_2'|<n^{7/8})\nonumber\\
    &\le \P\left(\text{Binom}\left(n,n^{-1/9}\right)\le n^{7/8}\right)\label{eq:lemmaE5-4}\\
    &\le \exp\left(-nD_{\mathrm{KL}}(n^{-1/8}||n^{-1/9})\right)\label{eq:lemmaE5-5}\\
    &=\exp\left(-n\cdot n^{-1/8}\log \left(n^{-1/72}\right)\right)\nonumber\\
    &\;\;\;\cdot \exp\left(-n\left(1-n^{-1/8}\right)\log\frac{1-n^{-1/8}}{1-n^{-1/9}}\right)\nonumber\\
    &=\exp\left(\frac{1}{72}n^{7/8}\log n-(1+o(1))n^{8/9}\right)\nonumber\\
    &=o(1)\nonumber,
\end{align}
where~\eqref{eq:lemmaE5-4} follows from~\eqref{eq:lemmaE5-3} and~\eqref{eq:lemmaE5-5} follows from Lemma~\ref{lem:add_chernoff}.
\subsection{Proof of Lemma~\ref{lem:KLdiv}}

\underline{(1) \textbf{Proof for equation \eqref{eq6-1}}}: 
From the definition of the KL-divergence between {$ \Bern\left(\frac{\Delta}{\log{1/\theta}}\theta s^2\right)$ and $ \Bern(\theta s^2)$}, we have 
\begin{align}
\nonumber
    &D_\mathrm{KL}\left(\tfrac{\Delta}{\log{1/\theta}}\theta s^2 \Big| \Big| \theta s^2\right)\\
    \nonumber
    & = \tfrac{\Delta}{\log{1/\theta}}\theta s^2 \log\left(\tfrac{\Delta}{\log{1/\theta}}\right)\\  
    &\nonumber 
    \hspace{.5em} 
    + \left[1-\left(\tfrac{\Delta}{\log{1/\theta}}\right) \theta s^2\right] \log\left( \tfrac{1-(\frac{\Delta}{\log{1/\theta}}) \theta s^2}{1-\theta s^2 }\right)\\
    \label{eq6-3}
    &= \tfrac{\Delta}{\log{1/\theta}} \theta s^2 \log \left(\tfrac{\Delta}{\log{1/\theta}}\right)\nonumber\\
    & \hspace{.5em}
    +\left[1-\left(\tfrac{\Delta}{\log{1/\theta}}\right) \theta s^2\right] \left( \left(1-\tfrac{\Delta}{\log{1/\theta}} \right)\tfrac{\theta s^2}{1-\theta s^2} +o(\theta)\right)
\end{align}
Here equation \eqref{eq6-3} follows from the Taylor series for the logarithm $\log(1+x) = x +o(x)$ for $x=o(1)$. 
Within the terms from \eqref{eq6-3}, we have that $\Delta=\Theta(1)$ and $\theta=o(1)$, we thus get $\frac{\Delta}{\log{1/\theta}} = o(1)$ and  $\left(\frac{\Delta}{\log{1/\theta}}\right) \log\left(\frac{\Delta}{\log{1/\theta}}\right) = o(1)$. We therefore have
\begin{align}
    \nonumber
      &D_\mathrm{KL}\left(\frac{\Delta}{\log{1/\theta}}\theta s^2 \bigg| \bigg| \theta s^2\right)\\
     \nonumber
     &= o(\theta) + \left( \frac{\theta s^2}{1-\theta s^2} \right)\\
     \label{eq6-4}
     &=  o(\theta) +\theta s^2. 
\end{align}
Here, the last equality \eqref{eq6-4} follows because  $\theta=o(1)$.\\

\underline{(2) \textbf{Proof for equation \eqref{eq6-2}}}:
From the definition of the KL-divergence between {$ \Bern\left(\frac{\Delta}{\log{1/\theta}}\theta s^2\right)$ and $ \Bern(\theta^2 s^2)$}, we have 
\begin{align}
    \nonumber
    &D_\mathrm{KL}\left(\tfrac{\Delta}{\log{1/\theta}}\theta s^2 \Big| \Big| \theta^2s^2 \right) \\
    \nonumber
    & = \tfrac{\Delta}{\log{1/\theta}}\theta s^2 \log \left( \tfrac{\frac{\Delta}{\log{1/\theta}}}{\theta} \right)  \\
    \nonumber & \hspace{10pt} +\left[1-\left(\tfrac{\Delta}{\log{1/\theta}}\right) \theta s^2\right] \log\left( \tfrac{1-(\frac{\Delta}{\log{1/\theta}}) \theta s^2}{1-\theta^2 s^2 }\right)\\
    &= \left(\tfrac{\Delta}{\log{1/\theta}}\right)\theta s^2 \log \left( \tfrac{\frac{\Delta}{\log{1/\theta}}}{\theta} \right) \nonumber \\
    &  \label{eq6-5}
    \hspace{10pt} +\left[1-\left(\tfrac{\Delta}{\log{1/\theta}}\right) \theta s^2\right] \left(- \tfrac{(\frac{\Delta}{\log{1/\theta}})\theta s^2 - \theta^2 s^2}{1-\theta^2 s^2} +o(\theta)\right)\\
    \label{eq6-6}
    & =\left(\tfrac{\Delta}{\log{1/\theta}}\right) \theta s^2 \log \left( \tfrac{\frac{\Delta}{\log{1/\theta}}}{\theta} \right)  + {o(\theta)}\\
    \nonumber
    & = \Delta \theta s^2 \frac{\log\Delta+\log 1/\theta - \log \log 1/\theta}{\log 1/\theta} + o(\theta)\\
    \label{eq6-7}
    & = \Delta\theta s^2 + { o (\theta)}
\end{align}
Here \eqref{eq6-5} follows from the Taylor expansion of $\log(1+x)$. Equation \eqref{eq6-6} follows because $\frac{\Delta}{\log{1/\theta}}=o(1)$. 
Equation \eqref{eq6-7} follows because $\Delta=\Theta(1)$ and { $\log 1/\theta = \omega(1)$}.

\section{Detailed discussion in seeded alignment}
\label{sec:comparison-proof}
In this section, we provide further details on the discussion in the seeded graph alignment problem in Section~\ref{sec:comparison-seeded}. We prove Remark~\ref{rmk:compare_it} and Theorem~\ref{thm:compare} by comparing the feasible regions in Corollaries~\ref{cor:alg1} and~\ref{cor:alg2} to the information theoretic feasible region and the best known feasible region by efficient algorithms in literature.

\subsection{Comparison to the information-theoretic limits}
\label{subsec:compare_IT}
Let us first compare the feasible region of Algorithms \AttrRich{} and \AttrSparse{} to the information theoretic limit of seeded alignment. The next corollary follows readily from Theorem~\ref{thm:IT limits}.
\begin{corollary}[Information-theoretic limits on seeded graph alignment]
\label{coro-seedIT}
Consider the seeded \er\ pair model $\mathcal{G}(N, \alpha, p,s)$, with $1-p=\Theta(1)$, $s = \Theta(1)$, and  $(1-\alpha) N=\omega(1)$.\\
\textbf{Achievability: }
    In the regime where $p=O\left(\frac{1}{[\log( N(1-\alpha))]^2}\right)$, if
    \begin{align}
        Nps^2 \geq \log (N(1-\alpha)) +\omega(1), \label{eq:corIT_cond1}
    \end{align}
    then {exact alignment is achievable w.h.p.}\\
    In the regime where $p=\omega\left(\frac{1}{[\log (N(1-\alpha))]^2}\right)$, if 
    \begin{align}
        Nps^2- a_N \geq \log (N(1-\alpha)) +\omega(1), \label{eq:corIT_cond2}
    \end{align}
    {where   
    $a_N \triangleq \alpha N\left(\sqrt{ps^2(1-p+p(1-s)^2)}-ps(1-s)\right)^2-N \alpha p s^2
    =O(N\alpha p^{3/2}),$
then exact alignment is achievable w.h.p.}\\
    \textbf{Converse:} If
    \begin{align*}
        Nps^2 \leq \log (N(1-\alpha)) -\omega(1),
    \end{align*}
    then no algorithm achieves exact alignment w.h.p.
\end{corollary}

To compare the feasible region of Corollary~\ref{coro-seedIT} to the feasible regions of Corollaries~\ref{cor:alg1} and~\ref{cor:alg2}, we consider two different regimes for $p$. 
\begin{enumerate} \setlength{\itemsep}{1.5ex}

    \item 
    \underline{$p=O\left(\frac{1}{[\log (N(1-\alpha))]^2}\right)$.} In this regime, the feasible region in Corollary~\ref{coro-seedIT} is given by condition~\eqref{eq:corIT_cond1}: $Nps^2 \geq \log (N(1-\alpha)) +\omega(1)$. In Corollary~\ref{cor:alg1}, condition~\eqref{eq:cor1_cond2} requires that $Nps^2\ge (1+\epsilon)\log ((1-\alpha)N)$. Since we assume that $n=(1-\alpha)N=\omega(1)$, we can see that condition~\eqref{eq:cor1_cond2} implies condition~\eqref{eq:corIT_cond1}. Therefore, the feasible region in Corollary~\ref{cor:alg1} is a subset of the feasible region in Corollary~\ref{coro-seedIT}.
    
    \noindent Similarly, in Corollary~\ref{cor:alg2}, condition~\eqref{eq:cor2_cond2} can be written as $Nps^2\ge\frac{\log ((1-\alpha)N)+\omega(1)}{1-\alpha}$, which also implies condition~\eqref{eq:corIT_cond1} since $1-\alpha<1$. Therefore, the feasible region in Corollary~\ref{cor:alg2} is also a subset of the feasible region in Corollary~\ref{coro-seedIT}. 
    \item \underline{$p=\omega\left(\frac{1}{[\log (N(1-\alpha))]^2}\right)$.} In this regime, the feasible region in Corollary~\ref{coro-seedIT} is given by condition~\eqref{eq:corIT_cond2}: $Nps^2 \geq \log (N(1-\alpha))+ a_N +\omega(1)$. Since $s=\Theta(1)$ and $p=\omega\left(\frac{1}{[\log (N(1-\alpha))]^2}\right)$, it follows that $Nps^2=\omega(\log (N(1-\alpha))) $. Moreover, since $p=o(1)$ and $a_N=O(N\alpha p^{3/2})$, we have $Nps^2=\omega(a_N)$. Therefore, condition~\eqref{eq:corIT_cond2} is satisfied and thus Corollary~\ref{coro-seedIT} applies in this regime
    while Corollaries~\ref{cor:alg1} and~\ref{cor:alg2} requires further conditions to apply. Therefore, the feasible regions in Corollaries~\ref{cor:alg1} and~\ref{cor:alg2} are both subsets of the feasible region in Corollary~\ref{coro-seedIT}. 
\end{enumerate}
\subsection{Comparison to existing efficient algorithms}
\label{subsec:compare_efficient}
In this subsection, we prove Theorem~\ref{thm:compare} by thoroughly comparing the feasible region of the proposed algorithms in Corollaries~\ref{cor:alg1} and~\ref{cor:alg2} to that of the existing algorithms in Theorems~\ref{cor:jiamingthm4},~\ref{cor:jiamingthm3} and~\ref{thm:erkip_thm2}. We discuss three regimes for $m$.

     
     \vspace{.5em}
         \emph{\textbf{Case 1:}} $m=o(n)$. Recall that $\alpha\triangleq\frac{m}{m+n}$. Therefore, we have $\alpha=o(1)$ in this regime. For the discussion in this regime, we first show that both Corollaries~\ref{cor:alg1} and~\ref{cor:alg2} apply in this regime. 
    We then show that the feasible region in Corollaries~\ref{cor:alg1} and~\ref{cor:alg2} extends the feasible region for exact alignment in Theorems~\ref{cor:jiamingthm4},~\ref{cor:jiamingthm3} and~\ref{thm:erkip_thm2}.
    
 To see Corollary~\ref{cor:alg1} applies in this regime, note that when $\alpha=o(1)$, condition~\eqref{eq:cor1_cond1} $\alpha Nps^2=\Omega(\log((1-\alpha)N))$ implies condition~\eqref{eq:cor1_cond2} $Nps^2\ge(1+\epsilon)\log((1-\alpha)N)$. Therefore, the feasible region in Corollary~\ref{cor:alg1} simplifies to condition~\eqref{eq:cor1_cond1}, which is non-empty when $p$, $s$ and $\alpha$ are large enough. 
    To see Corollary~\ref{cor:alg2} applies in this regime, we notice that conditions~\eqref{eq:cor2_cond1} $\alpha Nps^2=o(\log( (1-\alpha) N))$ and~\eqref{eq:cor2_cond2} $(1-\alpha)Nps^2-\log( (1-\alpha)N)\ge\omega(1)$ together imply that $\alpha=o(1)$ which is exactly the case in this regime.
    
     Now we move on to compare
    the feasible region in Corollaries~\ref{cor:alg1} and~\ref{cor:alg2} to that in Theorems~\ref{cor:jiamingthm4},~\ref{cor:jiamingthm3} and~\ref{thm:erkip_thm2}. We consider three different regimes for $Np$. 
         
         \vspace{.5em}
        \emph{\textbf{Case 1.1:}} $Np\le N^{1/7}$. 
        For this regime of $Np$, both Theorems~\ref{cor:jiamingthm4} and~\ref{cor:jiamingthm3} apply.
        We will show that the feasible regions in Corollaries~\ref{cor:alg1} and~\ref{cor:alg2} are covered by the feasible region in Theorem~\ref{cor:jiamingthm3} to illustrate that the proposed algorithms do not introduce any new feasible region in this regime. In this and all later cases, we always set $\beta=1/7$ when we apply Theorem~\ref{cor:jiamingthm3}.
        
        We first argue that the feasible region in Corollary~\ref{cor:alg1} is a subset of that in Theorem~\ref{cor:jiamingthm3}. This is because condition~\eqref{eq:cor1_cond1} $\alpha=\Omega(\frac{\log((1-\alpha)N)}{Nps^2})$ implies condition~\eqref{eq:jiamingthm3_cond2} $\alpha\ge N^{-4/7}$ under the assumption that $\alpha=o(1)$ and $Np\le N^{1/7}$, and condition~\eqref{eq:cor1_cond2} $Nps^2\ge (1+\epsilon)\log ((1-\alpha)N)$ implies condition~\eqref{eq:jiamingthm3_cond1} $Nps^2\ge \log N+\omega(1)$ under the assumption that $\alpha=o(1)$.
        
        Now we move on to show that the feasible region in Corollary~\ref{cor:alg2} is also a subset of that in Theorem~\ref{cor:jiamingthm3}. This is because conditions~\eqref{eq:cor2_cond2} $(1-\alpha)Nps^2-\log( (1-\alpha)N)\ge\omega(1)$ and~\eqref{eq:cor2_cond4} $\alpha Nps^2\ge\frac{2\log ((1-\alpha) N)}{\tau\log \frac1p}$ together imply condition~\eqref{eq:jiamingthm3_cond2} $\alpha\ge N^{-4/7}$ under the assumption that $\alpha=o(1)$ and $Np\le N^{1/7}$, and condition~\eqref{eq:cor2_cond2} $(1-\alpha)Nps^2-\log( (1-\alpha)N)\ge\omega(1)$ implies condition~\eqref{eq:jiamingthm3_cond1} $Nps^2\ge \log N+\omega(1)$ under the assumption that $\alpha=o(1)$.

        \vspace{.5em}
    \emph{\textbf{Case 1.2:}} {$N^{1/7}<Np\le\frac{s}{16(2-s)^2}N^{1/2}$.}  In this case, both Theorems~\ref{cor:jiamingthm4} and~\ref{cor:jiamingthm3} apply. We will show that the feasible regions in Corollaries~\ref{cor:alg1} and~\ref{cor:alg2} are covered by the feasible region in Theorem~\ref{cor:jiamingthm4} to illustrate that the proposed algorithms do not introduce any new feasible region in this regime.

    
     Firstly, we argue that the feasible region in Corollary~\ref{cor:alg1} is a subset of that in Theorem~\ref{cor:jiamingthm4}. Note that because $N^{1/7}<Np\le\frac{s}{16(2-s)^2}N^{1/2}$, we can write $Np=bN^a$ for some $a$ and $b$ such that $a,b=\Theta(1)$ and 
    $0<a\le\tfrac{1}{2} \quad\text{and} \quad 0<b\le \tfrac{s}{16(2-s)^2}.$ 
    Because $a\le \frac12$, we have $\floor{1/a}\ge 2$. Then the statement follows because condition~\eqref{eq:cor1_cond1} $\alpha Nps^2=\Omega(\log((1-\alpha)N))$ implies condition~\eqref{eq:jiamingthm4_cond} $\alpha\ge\frac{300\log N}{(Nps^2)^{\floor{1/a}}}$ under the assumption that $\alpha=o(1)$.
    
    
    
    Now we move on to show that the feasible region in Corollary~\ref{cor:alg2} is also covered by the feasible region in Theorem~\ref{cor:jiamingthm4}. This is because again we have $\floor{1/a}\ge 2$ and it follows that conditions~\eqref{eq:cor2_cond2} $(1-\alpha)Nps^2-\log( (1-\alpha)N)\ge\omega(1)$ and~\eqref{eq:cor2_cond4} $\alpha Nps^2\ge\frac{2\log ((1-\alpha) N)}{\tau\log \frac1p}$ together imply condition~\eqref{eq:jiamingthm4_cond} $\alpha\ge\frac{300\log N}{(Nps^2)^{\floor{1/a}}}$ under that assumption that $\alpha=o(1)$.
    
    \vspace{.5em}
    \emph{\textbf{Case 1.3:}} {$Np>\frac{s}{16(2-s)^2}N^{1/2}$.} In this regime, Theorem~\ref{cor:jiamingthm3} does not apply since $Np=\Omega(N^{1/2})$. We compare the feasible regions in Corollaries~\ref{cor:alg1} and~\ref{cor:alg2} to those in Theorems~\ref{cor:jiamingthm4} and~\ref{thm:erkip_thm2} to show that the proposed algorithms \emph{strictly improve} the best known feasible region in this regime. 
    
    To compare with the existing results, we first consider the feasible region achieved by the proposed algorithms. Because we are in the dense regime $Np=\Omega(N^{1/2})$, conditions~\eqref{eq:cor1_cond2} $Nps^2\ge (1+\epsilon)\log ((1-\alpha)N)$ and~\eqref{eq:cor2_cond2} $(1-\alpha)Nps^2-\log( (1-\alpha)N)\ge\omega(1)$ are satisfied. Then conditions for Corollaries~\ref{cor:alg1} and~\ref{cor:alg2} reduce to constraints on $\alpha$.
    The constraint in Corollary~\ref{cor:alg1} is given by~\eqref{eq:cor1_cond1} $\alpha Nps^2=\Omega(\log((1-\alpha)N))$, and the constraint in Corollary~\ref{cor:alg2} is given by~\eqref{eq:cor2_cond1} $\alpha Nps^2=o(\log((1-\alpha)N))$ and~\eqref{eq:cor2_cond4}, which can be written as $\alpha=\Omega(\frac{\log((1-\alpha)N)}{Nps^2\log\frac{1}{p}})$.
    Taking the union of the two constraints on $\alpha$, the feasible region of the proposed algorithm simplifies to
    \begin{equation}
        \alpha=\Omega\left(\frac{\log ((1-\alpha)N)}{Nps^2\log\frac{1}{p}}\right).
        \label{eq:cor2+3}
    \end{equation}
    
    Now, we move on to consider the feasible region in Theorem~\ref{cor:jiamingthm4}. Because $Np>\frac{s}{16(2-s)^2}N^{1/2}$, if we write $Np=bN^a$ for some $a,b=\Theta(1)$ and $0<b\le \frac{s}{16(2-s)^2}$, we must have $\frac12<a< 1$. Therefore, we have $\floor{1/a}=1$ in this regime and the feasible region in Theorem~\ref{cor:jiamingthm4}
    reduces to 
    \begin{equation}
        \alpha\ge\frac{300\log N}{Nps^2}.
        \label{eq:jiaming4_eq}
    \end{equation}
    
    From the above argument, we see that if condition~\eqref{eq:cor2+3} is satisfied while condition~\eqref{eq:jiaming4_eq} and at least one of conditions~\eqref{eq:erkip_cond1} $\alpha=\omega\left(\frac{1}{NI(p,s)^2}\right)$ and~\eqref{eq:erkip_cond2} $\alpha\ge \frac{2\log N}{NI(p,s)}$ in Theorem~\ref{thm:erkip_thm2} are not, then the proposed algorithms achieve exact w.h.p. while none of the existing works~\cite{Mossel2020} and~\cite{Shirani2017} do. In the following, we further consider three sub-cases for $p$ to discuss which one of conditions~\eqref{eq:erkip_cond1},~\eqref{eq:erkip_cond2} and~\eqref{eq:jiaming4_eq} is the bottleneck condition in the existing works~\cite{Mossel2020} and~\cite{Shirani2017} and to see the range of $\alpha$ that is feasible by the proposed algorithms while not by~\cite{Mossel2020} and~\cite{Shirani2017}.
   
   \begin{itemize}[leftmargin=1em]
       \item 
    Suppose $p\log N\log\frac1p=\omega(1)$. In this sub-case, condition~\eqref{eq:erkip_cond2} $\alpha\ge \frac{2\log N}{NI(p,s)}$ implies condition~\eqref{eq:erkip_cond1} $\alpha=\omega(\frac{1}{NI^2(p,s)})$ and condition~\eqref{eq:jiaming4_eq} $\alpha\ge\frac{300\log N}{Nps^2}$ implies condition~\eqref{eq:erkip_cond2}. Then the bottleneck condition is condition~\eqref{eq:erkip_cond2}. Therefore, if condition~\eqref{eq:cor2+3} is satisfied and $\alpha<\frac{2\log N}{NI(p,s)}$, the proposed algorithms achieve exact alignment w.h.p while the algorithms in~\cite{Mossel2020} and~\cite{Shirani2017} do not. Such $\alpha$ exists because $I(p,s)=(1+o(1))s^2p\log \frac1p$ and thus $\frac{2\log N}{NI(p,s)}=\Theta(\frac{\log ((1-\alpha)N)}{Nps^2\log\frac{1}{p}})$.
    
    \item Suppose $p\log N\log\frac1p=O(1)$ and $p\log N\log^2\frac1p=\omega(1)$. In this sub-case, condition~\eqref{eq:erkip_cond1} $\alpha=\omega(\frac{1}{NI^2(p,s)})$ implies condition~\eqref{eq:erkip_cond2} $\alpha\ge \frac{2\log N}{NI(p,s)}$  and condition~\eqref{eq:jiaming4_eq} $\alpha\ge\frac{300\log N}{Nps^2}$ implies condition~\eqref{eq:erkip_cond1}. Then the bottleneck condition is condition~\eqref{eq:erkip_cond1}. Therefore, if condition~\eqref{eq:cor2+3} is satisfied and $\alpha=O(\frac{1}{NI^2(p,s)})$, the proposed algorithms achieve exact alignment w.h.p while the algorithms in~\cite{Mossel2020} and~\cite{Shirani2017} do not. Such $\alpha$ exists because $\frac{\log ((1-\alpha)N)}{Nps^2\log\frac{1}{p}}=O(\frac{1}{NI^2(p,s)})$ under the assumption that $p\log N\log\frac1p=O(1)$. 
    
    \item Suppose $p\log N\log^2\frac1p=O(1)$. In this sub-case, condition~\eqref{eq:erkip_cond1} $\alpha=\omega(\frac{1}{NI^2(p,s)})$ implies condition~\eqref{eq:erkip_cond2} $\alpha\ge \frac{2\log N}{NI(p,s)}$  and condition~\eqref{eq:erkip_cond1} implies condition~\eqref{eq:jiaming4_eq} $\alpha\ge\frac{300\log N}{Nps^2}$. Then the bottleneck condition is condition~\eqref{eq:jiaming4_eq}. Therefore, if condition~\eqref{eq:cor2+3} is satisfied and $\alpha<\frac{300\log N}{Nps^2}$, the proposed algorithms achieve exact alignment w.h.p while the algorithms in~\cite{Mossel2020} and~\cite{Shirani2017} do not. Such $\alpha$ exists because $\frac{\log ((1-\alpha)N)}{Nps^2\log\frac{1}{p}}=o(\frac{300\log N}{Nps^2})$.
    \end{itemize}

    To summarize, in the regime of $m=o(n)$, the proposed Algorithms \AttrRich{} and \AttrSparse{} together \emph{strictly improve} the best known feasible region for exact alignment when $Np>\frac{s}{16(2-s)^2}N^{1/2}$. In the other two regimes for $Np$, the feasible region of the proposed algorithms is a strict subset of that in~\cite{Mossel2020}.
      
      \vspace{.5em}
    \emph{\textbf{Case 2:}} $m=\Omega(n)$ and $m=O(n^{1+\epsilon'})$ for some positive constant $\epsilon'$ that satisfies $\epsilon'<\epsilon$ and $\epsilon-\epsilon'=\Theta(1)$
    (cf. the constant $\epsilon$ in Corollary~\ref{cor:alg1}). In this regime, we have $\alpha=\Theta(1)$. For the discussion in this regime, we first show that Corollary~\ref{cor:alg1} applies in this regime while Corollary~\ref{cor:alg2} does not. 
    We then show that  the feasible region in Corollary~\ref{cor:alg1} is a subset of the feasible regions in Theorems~\ref{cor:jiamingthm4} and~\ref{cor:jiamingthm3} from existing work~\cite{Mossel2020}.

    Since we have $\alpha=\Theta(1)$, it follows that $Nps^2\ge (1+\epsilon)\log ((1-\alpha)N)$ implies $\alpha Nps^2=\Omega(\log((1-\alpha)N))$, i.e., condition~\eqref{eq:cor1_cond2} implies~\eqref{eq:cor1_cond1}. Therefore, the feasible region in Corollary~\ref{cor:alg1} simplifies to condition~\eqref{eq:cor1_cond2} which holds true when $p$ and $s$ are large enough. 
    As we argued in Case 1, Corollary~\ref{cor:alg2} requires $\alpha=o(1)$, so Corollary~\ref{cor:alg2} does not apply in this regime.

     Now we move on to compare the feasible region in Corollary~\ref{cor:alg1} to that in Theorems~\ref{cor:jiamingthm4} and~\ref{cor:jiamingthm3}. We further consider two different regimes for $Np$.
    
    \vspace{.5em}
        \emph{\textbf{Case 2.1:}} $Np=\exp(o(\log N))$. We can see that Theorem~\ref{cor:jiamingthm3} applies in this regime since $Np=\exp(o(\log N))\le N^{1/7}$.
        Condition~\eqref{eq:jiamingthm3_cond2} $\alpha\ge N^{-1+3/7}$ is satisfied without any further assumptions since $\alpha=\Theta(1)$. Moreover, condition~\eqref{eq:cor1_cond2} $Nps^2\ge(1+\epsilon)\log((1-\alpha)N)$ implies condition~\eqref{eq:jiamingthm3_cond1} $Nps^2\ge \log N+\omega(1)$ under the assumption that $m=O(n^{1+\epsilon'})$ and thus the feasible region in Corollary~\ref{cor:alg1} is a subset of the feasible region in Theorem~\ref{cor:jiamingthm3}.

      \vspace{.5em}  
    \emph{\textbf{Case 2.2:}} $Np=\exp(\Omega(\log N))$. Since $Np=\exp(\Omega(\log N))$ and we assume that $p=o(1)$, we can write $Np=bN^a$, where $a,b=\Theta(1)$, $0<a\le 1$, and $0<b\le \frac{s}{16(2-s)^2}$. Therefore, Theorem~\ref{cor:jiamingthm4} applies in this regime. Furthermore, condition~\eqref{eq:jiamingthm4_cond} in Theorem~\ref{cor:jiamingthm4} is satisfied since $\alpha=\Theta(1)\ge \frac{300\log N}{(Nps^2)^{\floor{1/a}}}$. We can see that all the conditions in Theorem~\ref{cor:jiamingthm4} are satisfied in this regime without any further assumptions.
        
         For Corollary~\ref{cor:alg1}, we have $Nps^2=\exp(\Omega(\log N))\ge (1+\epsilon)\log((1-\alpha)N)$, i.e., condition~\eqref{eq:cor1_cond2} is satisfied. We can see that all conditions in Corollary~\ref{cor:alg1} are also satisfied in this regime without any further assumptions. 
    
    To summarize, in the regime $m=\Omega(n)$ and $=O(n^{1+\epsilon'})$, the feasible region of the proposed algorithms is a subset of the feasible region given by existing work~\cite{Mossel2020}.

    \vspace{.5em}

    \emph{\textbf{Case 3:}} $m=\Omega(n^{1+\epsilon})$. In this regime, we have $\alpha=1-o(1)$. For the same reason as in Case 2, Corollary~\ref{cor:alg1} applies while Corollary~\ref{cor:alg2} does not. 
    We will show that the feasible region in Corollary~\ref{cor:alg1} extends the best known region for exact alignment in Theorems~\ref{cor:jiamingthm4},~\ref{cor:jiamingthm3} and~\ref{thm:erkip_thm2}. We discuss two regimes for $Np$.
    \vspace{.5em}
        \emph{\textbf{Case 3.1:}} $Np=\exp(o(\log N))$. We first point out that Theorem~\ref{thm:erkip_thm2} does not apply in this regime. This is because $I(p,s)=(1+o(1))s^2p\log \frac1p$ and $p\log\frac1p=\frac{\exp(o(\log N))}{N}\log\frac{N}{\exp(o(\log N))}=o(\frac{1}{\sqrt{N}})$,
        so condition~\eqref{eq:erkip_cond1} $\alpha=\omega\left(\frac{1}{NI(p,s)^2}\right)$ cannot be satisfied.
        Therefore, we focus on showing that the feasible region in Corollary~\ref{cor:alg1} strictly improves in the feasible regions in Theorem~\ref{cor:jiamingthm3}. This is because $m=\Omega(n^{1+\epsilon})$ is equivalent as $N=\Omega(((1-\alpha)N)^{1+\epsilon})$, and it follows that condition~\eqref{eq:jiamingthm3_cond1} $Nps^2\ge\log N+\omega(1)$ implies condition~\eqref{eq:cor1_cond2} $Nps^2\ge(1+\epsilon)\log((1-\alpha)N)$. So in the case when $Nps^2-\log N=O(1)$ and $Nps^2\ge(1+\epsilon)\log((1-\alpha)N)$ the proposed Algorithm \AttrRich{} achieves exact alignment w.h.p., while the algorithms in~\cite{Mossel2020} does not.
        
        \vspace{.5em}
        \emph{\textbf{Case 3.2:}} $Np=\exp(\Omega(\log N))$. 
        This case is included in the feasible region of both Corollary~\ref{cor:alg1} and Theorem~\ref{cor:jiamingthm4}
        for the same reason as in Case 2. Theorem~\ref{thm:erkip_thm2} further requires conditions~\eqref{eq:erkip_cond1} and~\eqref{eq:erkip_cond2} to apply in this regime.

    To summarize, in the regime $m=\Omega(n^{1+\epsilon})$, the proposed Algorithm \AttrRich{} 
    \emph{strictly improves} the best known feasible region for exact alignment in the case when $Np=\exp(o(\log N))$. In the case when $Np=\exp(\Omega(\log N))$, the feasible region by the proposed algorithms is a subset of that in existing literature~\cite{Mossel2020}.

From the above discussion, when specialized to the seeded \er{} graph pair model, the feasible region in Corollaries~\ref{cor:alg1} and~\ref{cor:alg2} strictly improves the best known feasible region in~\cite{Mossel2020} and~\cite{Shirani2017}, as shown in the blue area in Fig.~\ref{fig:seeded-comparison}. We note, however, that there is also some region that is feasible by existing results but not feasible by the proposed algorithms in this paper, as shown in the red area in Fig.~\ref{fig:seeded-comparison}.
 
 \section*{Acknowledgment}
This work was supported in part by the NSERC Discovery Grant No.\ RGPIN-2019-05448, the NSERC Collaborative Research and Development Grant CRDPJ 54367619, and the NSF grant CNS-2007733.

\bibliographystyle{IEEEtranN}
\bibliography{bibliography}

\end{document}